\documentclass[twocolumn,showpacs,preprintnumbers,amsmath,amssymb,pre,superscriptaddress]{revtex4-1}

\usepackage{graphicx}
\usepackage{dcolumn}
\usepackage{overpic}
\usepackage{float}
\usepackage{hyperref}
\usepackage{framed, color}
\usepackage{bbm}
\usepackage[dvipsnames]{xcolor}

\usepackage{amssymb}
\usepackage{amsmath}
\usepackage{siunitx}

\newcommand*\diff{\mathop{}\!\mathrm{d}}

\newcommand{\op}{\widehat}
\newcommand{\ket}{\rangle}

\newcommand{\+}{\dagger}
\newcommand{\e}{\mathrm{e}}
\newcommand{\dd}{\mathrm{d}}

\newcommand{\minf}{\scriptscriptstyle{-\infty}}
\newcommand{\pinf}{\scriptscriptstyle{+\infty}}

\newcommand{\intzp}{\int_0^{\pinf}\hspace{-1em}}
\newcommand{\intmp}{\int_{\minf}^{\pinf}\hspace{-.7em}}

\newcommand{\vac}{\mathbf{0}}

\begin{document}

\title{A semiclassical analysis of dark state transient dynamics in waveguide circuit QED} 

\author{ E. Wiegand    }
\email{wiegand@chalmers.se}
\affiliation{Applied Quantum Physics Laboratory, Department of Microtechnology and Nanoscience - MC2, Chalmers University of Technology, 412 96 G\"oteborg, Sweden}

\author{ B. Rousseaux    }
\affiliation{Applied Quantum Physics Laboratory, Department of Microtechnology and Nanoscience - MC2, Chalmers University of Technology, 412 96 G\"oteborg, Sweden}
\affiliation{Department of Physics, Chalmers University of Technology, 412 96 G\"oteborg, Sweden}

\author{ G. Johansson    }
\affiliation{Applied Quantum Physics Laboratory, Department of Microtechnology and Nanoscience - MC2, Chalmers University of Technology, 412 96 G\"oteborg, Sweden}

\begin{abstract} 
	 The interaction between superconducting qubits and one-dimensional microwave transmission lines has been studied experimentally and theoretically in the past two decades. In this work, we investigate the spontaneous emission of an initially excited artificial atom which is capacitively coupled to a semi-infinite transmission line, shorted at one end. This configuration can be viewed as an atom in front of a mirror. The distance between the atom and the mirror introduces a time-delay in the system, which we take into account fully. When the delay time equals an integer number of atom oscillation periods, the atom converges into a dark state after an initial decay period. The dark state is an effect of destructive interference between the reflected part of the field and the part directly emitted by the atom. Based on circuit quantization, we derive linearized equations of motion for the system and use these for a semiclassical analysis of the transient dynamics. We also make a rigorous connection to the quantum optics system-reservoir approach and compare these two methods to describe the dynamics. We find that both approaches are equivalent for transmission lines with a low characteristic impedance, while they differ when this impedance is higher than the typical impedance of the superconducting artificial atom.
\end{abstract}


\maketitle

%
%
%
\section{Introduction}

Waveguide quantum electrodynamics (waveguide QED) has become a field of growing importance for quantum communication, quantum simulations and quantum computation \cite{,Roy2017, Gu2017,GonzCira17_2,PaulGonz18,QuijJoha18}. In waveguide QED, the interaction between a one-dimensional (1D) electromagnetic field and quantum emitters is studied \cite{ZhenBara10, Tufarelli2014,SancMart14}. The restriction to one dimension gives an advantage in transferring information, since it increases directionality and reduces losses \cite{SancZuec16,SancGarc17}. The quantum emitters can be natural atoms, Rydberg atoms, trapped ions or artificial atoms such as quantum dots, nitrogen vacancy centers and superconducting qubits \cite{Roy2017}. The latter are studied in a newer field called circuit quantum electrodynamics (circuit QED)
\cite{Wallraff2004,Blais2004,Wendin2017,Gu2017,Kockum2019}.

   In this field, superconducting circuits including Josephson junctions (JJs), work in the microwave regime. Like natural atoms, these circuits have a discrete and anharmonic energy spectrum and can therefore be used as qubits. Circuit-QED artificial atoms can thus mimic atom and molecular dynamics at the quantum level. Furthermore they enable the exploration of new parameter regimes such as reaching the strong and ultrastrong coupling regimes, where light and matter are no longer separable \cite{DevoScho07,BambOgaw14,USC_Anton,ZuecGarc19,MlynWall14}, or opening the possibility of observing the superradiant phase transition \cite{NataCiut10,BambNaka16,BambNobu17}.
   
 Hoi \emph{et al} \cite{Hoi2015} coupled a so-called transmon \cite{Koch2007} qubit to a one-dimensional microwave transmission line (TL) which was shortened at one end. A transmon qubit is most easily understood as an LC oscillator whose inductance is made nonlinear with a JJ. This system is usually described as an atom in front of a mirror \cite{Wen2018,Wen2019,Peng2016,FornDiaz2017}, since the microwaves are reflected at the shorted end of the TL (the mirror) and interact with the qubit again. The effective distance of the qubit to the mirror with respect to the wavelength of the field plays a crucial role for the dynamics of the system. Hoi \emph{et al} showed that the qubit can be hidden if it is placed at a node of the field, meaning that it does not interact with the field and the spontaneous emission rate vanishes \cite{Hoi2015}. This was shown theoretically by using a master equation approach with a Markov approximation. In the experiment the atom was probed in reflection, and a suppression of the spontaneous emission rate with a factor of 50 compared to the open TL case was verified.
 
However, considering an initially excited atom and a vacuum state in the TL, there is a time $T$ given by the velocity of light and the distance to the mirror and back, during which the atom will decay with a rate $\gamma$ given by the open TL case. The reduction of the decay rate corresponds to a destructive interference between the light emitted from the atom and the reflected light from the mirror. To resolve the dynamics on this timescale, one needs to go beyond the Markov approximation, including effects of the time-delay beyond phase-shifts.\\
This has been done in several studies investigating light-matter interaction regarding time delay, such as quantum optical approaches solving the equations of motion with Fourier transformation \cite{Dorner2002,Guo2017, Tufarelli2014}, recent methods involving matrix product states to solve time-delay equations \cite{Grimsmo2015,Pichler2016,Pichler2017,Guimond_2017}, or Green's function approaches \cite{SchnBusc16,GonzCira17}. However, these all rely on a weak-coupling approximation between the atom and the waveguide, where one degree of freedom of the atom couples to the transmission line in one point.

In this paper, we investigate the spontaneous emission rate of an initially excited transmon qubit which can be placed at an arbitrary distance to the mirror. For long distances, $\gamma T \sim 1$, we take time-delay effects into account, i.e., we go beyond the Markov approximation of Ref.~\cite{Hoi2015}. Using circuit quantization, we derive equations of motion in principle valid beyond the weak coupling regime.

In section \ref{circuitQED}, we derive the circuit-QED equations of a single transmon capacitively coupled to a TL and describe its decay dynamics in different regimes. In section \ref{sysres}, we derive a rigorous connection between the circuit QED and the system-reservoir approaches. We then compare the transient dynamics in the two models and discuss the applicability of the system-reservoir approach for this system. In section \ref{sec:Wiggles}, we discuss the fast oscillating terms of the semi-classical model. Then finally, in section \ref{conclusion} we summarize the results and discussions presented in this article.
%
%
%
%
%
 \section{Circuit-QED model}
 \label{circuitQED}
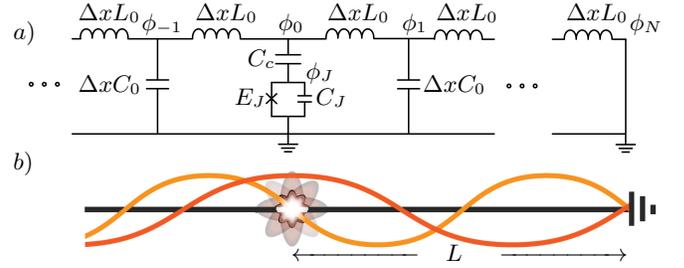
\begin{figure}[t]
	
	\centering
\begin{overpic}[width=1\linewidth]{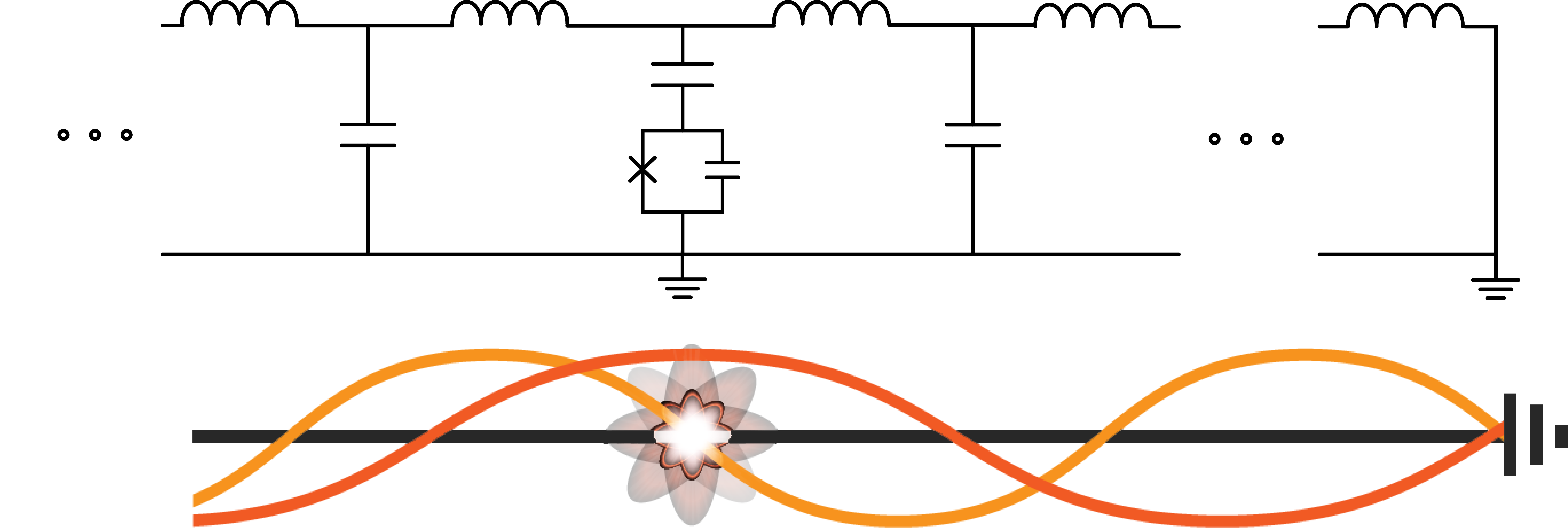}
	\put(21,33.5){$\phi_{-1}$}
	\put(42,33.5){$\phi_{0}$}
	\put(61,33.5){$\phi_{1}$}
	\put(96,33.5){$\phi_{N}$}
	\put(36,28){ $C_c$}
	\put(45,26){ $\phi_J$}
	\put(34,22){ $E_J$}
	\put(10,35){ $\Delta x L_0$}
	\put(28,35){ $\Delta x L_0$}
	\put(48.3,35){ $\Delta x L_0$}
	\put(65,35){ $\Delta x L_0$}
	\put(85,35){ $\Delta x L_0$}
	\put(10,24){ $\Delta x C_0$}
	\put(63,24){ $\Delta x C_0$}
	\put(46.5,22){ $C_J$}
	\put(0,32){ $a)$}
	\put(0,12){ $b)$}
	\put(44,-2){$\xleftarrow{\hspace*{1.5cm}}$}
	\put(76,-2){$\xrightarrow{\hspace*{1.5cm}}$}
	\put(66.5,-2){ $L$}
		
	\end{overpic}
	\caption{a) A transmon is coupled to the 1D-TL by a coupling capacitance $C_c$. The TL is grounded at one end. The energy, flux and capacitance of the transmon are denoted by $E_J$, $\phi_J$ and $C_J$, respectively. The TL is modelled by coupled LC oscillators with capacitance $\Delta x C_0$ and inductance $\Delta x L_0$. The flux of the nodes between the LC oscillators are denoted by $\phi_n$.
		b) Sketch of the system, depicting an atom in front of a mirror. The atom can be coupled (decoupled) to the electro-magnetic field depending on its location at the antinode (node) of an electro-magnetic mode. The distance between the atom and the mirror is denoted as $L$.}
	\label{fig:System}
\end{figure}

Our system consists of a transmon, capacitively coupled to an open 1D-TL, which is grounded at one end. A transmon is a superconducting qubit, that consists of a JJ with Josephson energy $E_J$ and a capacitance $C_J$ in parallel. The nonlinearity of the JJ yields an anharmonic excitation spectrum for the transmon. The TL is a one-dimensional coplanar waveguide, with a characteristic inductance/capacitance $L_0/C_0$ per unit length. It supports a TEM mode with microwaves propagating at the velocity $v_0=1/\sqrt{L_0 C_0}$.  The TL is modeled as a discretized circuit consisting of coupled LC oscillators, using a discretization length $\Delta x$, much shorter than the wavelength of the microwaves. Our semi-infinite TL is shorted at one end, where the electromagnetic field is reflected. The transmon is coupled to the TL by a capacitance $C_c$ at a distance $L$ from the shorted end. In the discretized model, we number the transmission line node coupled to the transmon as node zero. We then ground node $N=L/\Delta x$ to the right of the transmon ($\phi_N=0$). A sketch of the circuit model and the system is depicted in Fig. \ref{fig:System} a). Due to the shorted TL, the system can be described as an atom in front of a mirror~\cite{Hoi2015}, see Fig \ref{fig:System} b). When the qubit is excited and decays, it emits electromagnetic excitations into the TL, which initially start to propagate in both directions. The light propagating to the left is lost while the light propagating to the right is reflected at the mirror. The reflected light interacts with the qubit again after a time delay $T=2L/v_0$, given simply by the distance to the mirror and the velocity of light in the TL. 
To describe our circuit we use the node fluxes $\phi_i (t) = \int_{0}^{t} V_i (t') dt'$ as coordinates, where $V_i(t)$ are the voltages at the node $i$. Using the circuit quantization procedure of Ref.~\cite{Devoret1995}, the Hamiltonian corresponding to the system is given by
\begin{align}
\label{Hphip}
	H(\phi_i,p_i) = \frac{1}{2C_0\Delta x}\sum_{n \neq 0} p_{n \neq 0}^2(t)\nonumber \\ 
	+ \frac{1}{2C_c} p_0^2(t) + \frac{1}{2 C_J}\left(p_J(t) + p_0(t)\right)^2 \nonumber \\
	+ \sum_{n=-\infty}^{N-1} \frac{1}{2L_0\Delta x} \left( \phi_{n+1}(t) -  
	\phi_n (t) \right)^2 \nonumber \\ - E_J \cos(\frac{2e}{\hbar}\phi_J(t)),
\end{align}
where the charges $p_i$ are the conjugate momenta of the node fluxes, fulfilling the canonical commutation relations, $\left[ \phi_i, p_j\right] = i \hbar \delta_{ij}, \left[\phi_i, \phi_j\right] = \left[p_i, p_j\right] = 0$. 
%
%

From the Heisenberg equations of motion for an operator $A(t)$ by $\frac{d}{dt} A(t) = \frac{i}{\hbar} \left[H, A(t)\right]$, we can now derive the coupled equations of motion for all our operators
\vskip0.3cm
\begin{align}
	\partial_t \phi_J (t) &= \frac{1}{C_J} (p_J(t)  + p_0(t) ), \label{eq:EoMphiJa}\\
	\partial_t p_J (t) &= - E_J \frac{2 e}{\hbar} \sin ( \frac{2 e}{\hbar}\phi_J(t) ) , \label{eq:EoMpJa} \\
    \partial_t \phi_0 (t) &= \frac{C_c + C_J}{C_c C_J} p_0(t) + \frac{1}{C_J} p_J(t) , \label{eq:EoMphi0a}\\
    \partial_t \phi_i (t) &=\frac{p_i}{\Delta x C_0}, (i\neq 0), \label{eq:EoMpi}\\
    \partial_t p_i (t) &=\frac { 1 } { L _ { 0 } \Delta x } \left( \phi _ { i + 1 } ( t ) - 2 \phi _ { i } ( t )  + \phi _ { i - 1 } ( t ) \right). \label{eq:EoMp0a}
\end{align}

\subsection{The continuum limit}
We now take the continuum limit $\Delta x \rightarrow 0$ and replace the node fluxes $\phi_i(t)$ in the TL with a continuous flux field $\phi(x,t)$. We choose the coordinate $x_i=i \Delta x$, so that the transmon is now located at $x=0$. The TL charges $p_i(t)$ for $i\neq 0$ are replaced by a charge density field $p(x_i,t)=p_i(t)/\Delta x$ with dimension charge per unit length. This can be understood from the fact that the TL node charge vanishes together with the node capacitance $\Delta x C_0$, with the finite ratio $p_i(t)/\Delta x$. Away from the transmon, i.e., for $x\neq 0$, the equations of motion Eq.~(\ref{eq:EoMpi}) and Eq.~(\ref{eq:EoMp0a}) are replaced by 
\begin{align}
    \partial_t \phi(x,t) &=\frac{p(x,t)}{C_0},\\
    \partial_t p(x,t) &=\frac{\partial^2_x\phi(x,t)}{L_0}. \label{eq:EoMp0aCont}
\end{align}
These equations can be recognized as the massless Klein-Gordon equations in one spatial dimension.

\subsubsection{The free TL field}
We write the field in terms of bosonic creation and annihilation operators for plane waves  $a_k$ and $a_k^\dagger$ with wavenumber $k$, which obey the canonical commutation relations, $\left[a_k, a_{k'}^\dagger\right] = \delta (k-k')$ and $\left[ a_k,a_{k'} \right] = \left[ a_k^\dagger,a_{k'}^\dagger \right] = 0$,
\begin{align}
	\phi^\rightleftarrows (x,t) = \sqrt{\frac{\hbar}{4\pi C_0}} \int_{-\infty}^{\infty} \frac{\diff k}{\sqrt{\omega_k}} \left( a^\rightleftarrows_k e^{-i (\omega_k t \mp k x)} + \text{h.c.} \right).
	\label{eq:phifield}
\end{align}
Here, the arrows indicate right ($\rightarrow$) and left ($\leftarrow$) moving parts of the field, moving at the speed of light in the transmission line $v_0$. The corresponding expression for the charge density field $p(x,t)$ reads
\begin{align}
	p^\rightleftarrows (x,t) = i \sqrt{\frac{\hbar C_0}{4\pi}} \int_{-\infty}^{\infty} \diff k\sqrt{\omega_k} \left( a^\rightleftarrows_k e^{-i (\omega_k t \mp k x)} - \text{h.c.} \right).
\end{align}
We now rewrite Eq.~\eqref{eq:phifield} in terms of frequencies $\omega_k=v_0 |k|$ instead of wavenumbers $k$ and obtain
\begin{align}
	\phi^\rightleftarrows (x,t) = \sqrt{\frac{\hbar Z_0}{4\pi}} \int_{0}^{\infty} \frac{\diff \omega}{\sqrt{\omega}} \left( a^\rightleftarrows_\omega e^{-i (\omega t \mp k_\omega x)} + \text{h.c.} \right).
\end{align}
The voltage in the TL is given by the time derivative of the flux field, $V (x,t)= \partial_t \phi (x,t)$,
\begin{align}
	V^\rightleftarrows (x,t) = - i \sqrt{\frac{\hbar Z_0}{4\pi}} \int_{0}^{\infty} \diff \omega\sqrt{\omega} \left( a^\rightleftarrows_\omega e^{-i (\omega t \mp k_\omega x)} - \text{h.c.} \right),
\end{align}
while the current is proportional to the spatial derivative of the flux field, $I (x,t) = \partial_x \phi (x)/L_0$,
\begin{align}
	I^\rightleftarrows (x,t) = - i \sqrt{\frac{\hbar}{4\pi Z_0}} \int_{0}^{\infty} \diff \omega\sqrt{\omega}\left( a^\rightleftarrows_\omega e^{-i (\omega t \mp k_\omega x)} - \text{h.c.} \right).
\end{align}
\subsubsection{Scattering at the transmon}
We now want to connect the field in the TL to the transmon degrees of freedom at the point $x=0$. The flux field is continuous, so we can straightforwardly make the identification $\phi_0(t)=\phi(0,t)$. However, since the node $i=0$ has a finite capacitance also for $\Delta x\rightarrow 0$, we find that the node charge $p_0(t)$ remains finite and we need to keep that as a separate variable. This also implies that the spatial derivative of the flux field does not have to be continuous at $x=0$. Keeping this in mind in taking the continuum limit of Eq.~(\ref{eq:EoMp0a}) at $x=0$, we arrive at
\begin{equation}
 \partial_t p_0 (t) = \frac{1}{L_0}\left(\partial_x \phi(0^+,t)-\partial_x \phi(0^-,t) \right).  \label{eq:EoMp0aCont} 
\end{equation}
Using the continuity of the voltage at this point we obtain
\begin{equation}
	V_0 = \partial_t \phi_0 (t) =\partial_t \phi(0,t) = V_L^{\text{in}} + V_L^{\text{out}} = V_R^{\text{in}} + V_R^{\text{out}},
	\label{eq:delphi0}
\end{equation}
where $V_L^{\text{in}}=V^{\rightarrow}(0^-,t)$, $V_L^{\text{out}}=V^{\leftarrow}(0^-,t)$,  $V_R^{\text{in}}=V^{\leftarrow}(0^+,t)$ and $V_R^{\text{out}}=V^{\rightarrow}(0^+,t)$ are the in- and out-going voltage fields at the left (L) and right (R) side of the coupling point, respectively.
Furthermore, current conservation at this point in the circuit is expressed through Eq.~(\ref{eq:EoMp0aCont}) as
\begin{align}
	\partial_t p_0 &= \frac{1}{Z_0} \left( V_{\text{in}} - V_{\text{out}} \right) \nonumber \\
	&= \frac{1}{Z_0} \left( V_L^{\text{in}} + V_R^{\text{in}} - V_L^{\text{out}} - V_R^{\text{out}}  \right).
	\label{eq:delp0}
\end{align}
Combining Eqs.(\ref{eq:EoMphi0a}), (\ref{eq:delphi0}) and (\ref{eq:delp0}), we can eliminate $\phi_0(t)$ and obtain
\begin{equation}
    \frac{C_c + C_J}{C_c C_J} p_0(t) + \frac{1}{C_J} p_J(t) + \frac{Z_0}{2} \partial_t p_0(t) = V_L^{\text{in}}(t)+V_R^{\text{in}}(t), \label{Eq:EoMp0OpenTL}
\end{equation}
which together with Eqs. (\ref{eq:EoMphiJa}) and (\ref{eq:EoMpJa})
\begin{align*}
	\partial_t \phi_J (t) &= \frac{1}{C_J} (p_J(t)  + p_0(t) ), \\
	\partial_t p_J (t) &= - E_J \frac{2 e}{\hbar} \sin ( \frac{2 e}{\hbar}\phi_J(t) ),
\end{align*}
determines the transmon dynamics in terms of the incoming fields $V_L^{\text{in}}(t)$ and $V_R^{\text{in}}(t)$. 
From Eqs.(\ref{eq:EoMphi0a}), (\ref{eq:delphi0}) and (\ref{eq:delp0}) we also obtain the expressions for the outgoing fields
\begin{align}
	V_L^{\text{out}}(t) &= V_R^{\text{in}}(t) - \frac{Z_0}{2} \partial_t p_0(t), \\
	V_R^{\text{out}}(t) &= V_L^{\text{in}}(t) - \frac{Z_0}{2} \partial_t p_0(t).
\end{align}
Thus, we have derived the equations of motion for a transmon capacitively coupled to a TL.

\subsubsection{The mirror}
The field going away from the transmon to the right $V_R^{\text{out}}$ is reflected at the mirror and returns as $V_R^{\text{in}}$ with a time delay $T=2L/v_0$ and a $\pi$ phase shift acquired at the shorted mirror. 
We note that an open ended TL would result in the same time-delay, but no extra phase shift at the mirror. Thus eliminating $V_R^{\text{in}}$ we modify Eq.~(\ref{Eq:EoMp0OpenTL}) into
\begin{align}
    \frac{C_c + C_J}{C_c C_J} p_0(t) + \frac{1}{C_J} p_J(t) &+ \frac{Z_0}{2} \partial_t \left(p_0(t)\mp p_0(t-T)\right) \nonumber \\ &= V_L^{\text{in}}(t)\mp V_L^{\text{in}}(t-T), \label{Eq:EoMp0Mirror}
\end{align}
where the lower positive sign correspond to the open-ended mirror case.

This is a time-delay differential equation for the system operators, which together with the non-linearity of equation (\ref{eq:EoMpJa}) makes it impossible to find the general solution. However, in the next section we linearize the transmon qubit, considering the single-excitation regime, yielding analytically solvable equations of motion.

\subsection{Linearization of the transmon qubit}
In the weak coupling regime (specified in detail below) and neglecting the time delay, the system behaves as an atom coupled to a bath, where the coupling strength depends strongly on the distance to the mirror. 
Here we lay the foundations for exploring this system beyond the weak coupling regime, including the effects of time delay. Due to the limited anharmonicity of the transmon, a relevant approximation is then to neglect the non-linearity of the JJ and replace Eq.~(\ref{eq:EoMpJa}) with its linearized version
\begin{equation}
    \partial_t p_J(t)=-\frac{\phi_J(t)}{L_J},
    \label{Eq:pj_linearized}
\end{equation}
where we introduced the Josephson inductance
\begin{equation}
    L_J =\frac{\hbar^2}{4e^2E_J},
\end{equation}
by expanding the sine function to first order. This approximation is obviously good in the weak excitation regime $|\phi_J(t)| < \hbar/2e$. This leaves us with linear time-delay differential equations that we will explore in the rest of this paper. One property of linear quantum equations of motion is that the quantum averages can be taken directly and the average of the observables thus obey identical real-valued classical equations of motion. In particular, we will use this correspondence to explore the decay dynamics of an initially excited transmon.  

\subsection{An effective lumped element electrical circuit for the open TL case}
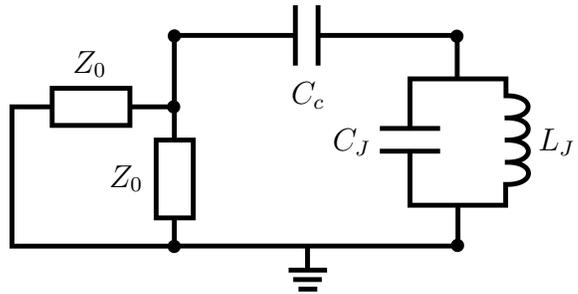
\begin{figure}[t]
	\centering
	\begin{overpic}[width=0.8\linewidth]{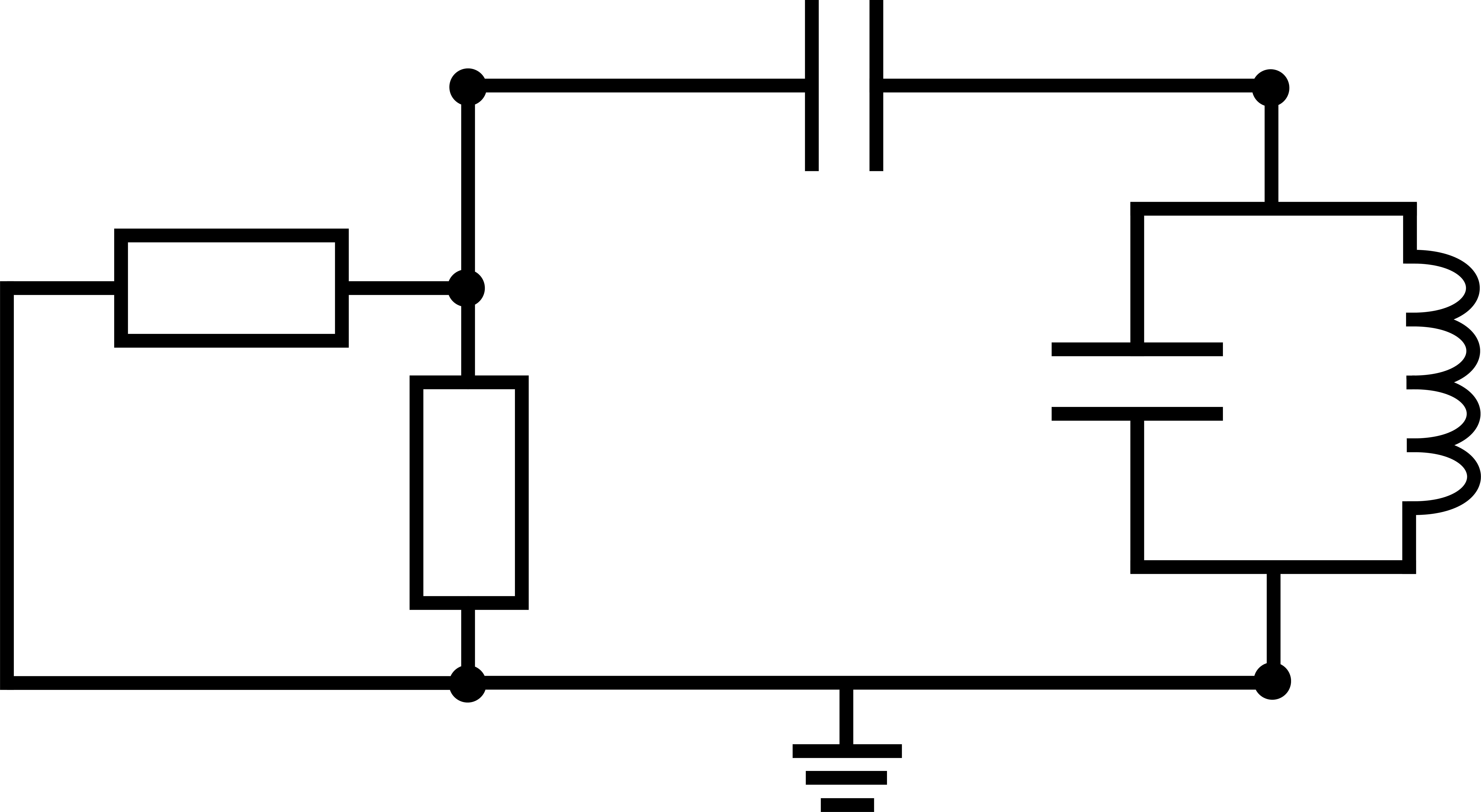}
		\put(62,27){\large $C_J$}
		\put(101.5,27){\large$L_J$}
		\put(54,36){\large$C_c$}
		\put(12,42){\large$Z_0$}
		\put(19,20){\large$Z_0$}
	\end{overpic}
	\caption{Simplified system of a transmon coupled to an open TL. The transmon corresponds to an LC oscillator with inductance $L_J$ and capacitance $C_J$. It is coupled to the TL with characteristic impedance $Z_0$ through the coupling capacitance $C_c$. Considering an open TL, the photon can escape in both directions, corresponding to the two resistors in parallel.}
	\label{fig:Effective_circuit}
\end{figure}

Having linearized the transmon, we will now analyze the coupling strength between the transmon and transmission line by studying the energy decay rate of an intially excited transmon to an open TL (no mirror), depending on the circuit parameters. We therefore also assume that there are no average fields incoming towards the transmon, i.e. $\langle V_L^{\text{in}}(t) \rangle = \langle V_R^{\text{in}}(t) \rangle=0$. The average charges $\bar{p}_J(t)=\langle p_J(t) \rangle$ and $\bar{p}_0(t)=\langle p_0(t) \rangle$ then obey the averaged versions of Eqs. (\ref{Eq:EoMp0OpenTL}), (\ref{eq:EoMphiJa}) and (\ref{Eq:pj_linearized}),
\begin{align}
    \frac{C_c + C_J}{C_c C_J} \bar{p}_0(t) &+ \frac{1}{C_J} \bar{p}_J(t) + \frac{Z_0}{2} \partial_t \bar{p}_0(t) = 0 \label{Eq:semiclassP0}\\
	\partial_t \bar{\phi}_J (t) &= \frac{1}{C_J} (\bar{p}_J(t)  + \bar{p}_0(t) ), \label{Eq:semiclassPhiJ} \\
	\partial_t \bar{p}_J (t) &= -\frac{\bar{\phi}_J(t)}{L_J}.
\label{Eq:semiclassPJ}
\end{align}
This undriven linearized transmon dynamics corresponds to the effective lumped element circuit in Fig.~\ref{fig:Effective_circuit}, which we can use to discuss the different parameter regimes more intuitively.

The only dissipative element in this circuit is the characteristic impedance $Z_0$ of the transmission line. If we set this to zero, Eq.~(\ref{Eq:semiclassP0}) leads to 
\begin{equation}
Z_0=0 \quad \Rightarrow \quad \bar{p}_0(t)=-\frac{C_c}{C_J+C_c}\bar{p}_J(t).
\label{Eq:P0forZ0eq0}
\end{equation}
This corresponds to an undamped harmonic LC-oscillator with angular frequency $\omega_0=1/\sqrt{L_J(C_c+C_J)}$, given by the two capacitances $C_J$ and $C_c$ connected in parallel to ground. Here we also note that the energy of the oscillator is given by
\begin{equation}
Z_0=0 \quad \Rightarrow \quad E_q=\frac{\bar{p}_J(t)^2}{2 (C_J + C_c)} + \frac{\bar{\phi}_J(t)^2}{2L_J}.
\label{Eq:EqforZ0eq0}
\end{equation}
If we instead set $Z_0$ to infinity, $C_c$ is connected to an open circuit 
\begin{equation}
Z_0 \rightarrow \infty  \quad \Rightarrow \quad \partial_t\bar{p}_0(t)=0
\label{Eq:P0forZ0eqInfinity}
\end{equation}
and we again find an undamped LC-oscillator, now with frequency $\omega_J=1/\sqrt{L_J C_J}$. 

For finite damping, it is useful to find expressions for the relaxation rate. We do this analysis by replacing the JJ with an ac current source of amplitude $i_J$ and angular frequency $\omega$. Using the Phasor method, we find that the average power dissipated in the transmission line is
\begin{equation}
    P_{Z_0}= \frac{i_J^2 C_c^2 Z_0}{4\left(C_J+C_c\right)^2+C_c^2 C_J^2 Z_0^2 \omega^2}.
\end{equation}
We also find the reactive ac power of the circuit, i.e., the average rate of energy the current source has to supply and reabsorb during a period
\begin{equation}
    P_{r}= \frac{i_J^2}{2\omega}\frac{4 \left(C_J+C_c\right) + C_c^2 C_J Z_0^2 \omega^2}{4\left(C_J+C_c\right)^2+C_c^2 C_J^2 Z_0^2 \omega^2}.
\end{equation}
Without dissipation, the energy stored in the oscillator/qubit would be given by
\begin{equation}
E_{q}=\frac{P_r}{\omega}.    
\end{equation}
In the weakly damped regime, corresponding to an atom weakly coupled to the field, the energy of the oscillator/atom decays exponentially $E_{q} (t) = E_{q}(0) e^{-\gamma t}$ and we now find an expression for the decay rate through
\begin{equation}
    \gamma=\frac{P_{Z_0}}{E_{q}}=\omega\frac{P_{Z_0}}{P_r}=\frac{2}{Z_0C_J}\frac{\eta}{1+\eta},
\end{equation}
where we defined the dimensionless parameter
\begin{equation}
    \eta=\omega^2\frac{Z^2_0C_c^2}{4}\frac{C_J}{C_J+C_c}.
\end{equation}
As mentioned, this estimation of the decay rate is relevant in the weak coupling regime, $\gamma/\omega < 1$. Using the approximation $\omega=1/\sqrt{L_J C_J}$, we find for this ratio
\begin{equation}
    \frac{\gamma}{\omega}=\frac{P_{Z_0}}{P_r}=2\frac{\sqrt{L_J/C_J}}{Z_0}\frac{\eta}{1+\eta}=2\frac{Z_J}{Z_0}\frac{\eta}{1+\eta},
    \label{FullOpenTLDecayRate}
\end{equation}
where in the last step we defined the qubit impedance $Z_J=\sqrt{L_J/C_J}$. Using the expression for the charging energy of the JJ, $E_C = e^2/(2C_J)$ and the resistance quantum $R_K=h/e^2 \approx 25 \, \si{k \Omega}$ , we can also write 
\begin{equation}
Z_J=\frac{R_K}{2\pi \sqrt{2}} \sqrt{\frac{E_C}{E_J}},
\end{equation}
to see that the qubit impedance is directly determined by the $E_J/E_C$-ratio. This ratio should be much larger than one, for the circuit to be in the charge-noise insensitive transmon regime.

In the regime of a low-impedance TL, characterized by $\eta < 1$, we expand the decay rate to first order in $\eta$ and using that the oscillator frequency in this regime is given by $\omega\approx 1/\sqrt{L_J(C_J+C_c)}$ we find
\begin{equation}
    \gamma\approx \frac{2\eta}{Z_0C_J}=\frac{Z_0}{2}\omega^2\frac{C_c^2}{C_J+C_c}\approx\frac{Z_0}{2 L_J}\frac{C_c^2}{\left(C_J+C_c\right)^2}.
    \label{Eq:gammaLowEta}
\end{equation}
Here, we note that $\eta < 1$ has been the relevant regime for all experiments using transmons and TLs of around $Z_0 = 50-100\, \si{\Omega}$ so far. In the experiment of Ref.~\cite{Hoi2015} we have, e.g., $\eta = 2.2\cdot 10^{-4}$.

Using a TL with inductances made from Josephson junctions or high kinetic inductance materials, it is possible to reach characteristic impedances of a few k$\si{\Omega}$ \cite{Weissl2015,Krupko2018,Masluk2012}. This would be necessary to approach the regime $\eta \sim 1$, where the largest coupling ratio $\gamma/\omega=Z_J/4Z_0$ would be obtained according to this simple analysis.

\subsection{Spontaneous emission in front of a mirror}
We now return to the transmon in front of a mirror to study the effect of the time delay $T$ caused by the finite distance to the mirror. To study the spontaneous emission, we again look at the classical linearized equation of motion for the averaged observables, with no incoming field
\begin{equation}
	\frac{C_c + C_J}{C_c C_J} p_0(t) + \frac{1}{C_J} p_J(t) = - \frac{Z_0}{2} \partial_t \left(  p_0 (t) -  p_0 (t - T) \right), \label{eq:EoMredpJ1}\\
\end{equation}
which we obtain by performing a quantum average of Eq.~(\ref{Eq:EoMp0Mirror}) with a shorted mirror. To simplify the notation in the following, we use the symbols $p_0(t)$ and $p_J(t)$ also for the averaged observables.
Combining Eqs.~(\ref{Eq:semiclassPhiJ}) and (\ref{Eq:semiclassPJ}) into 
\begin{equation}
	\partial_t^2 p_J (t) = - \omega_J^2 \left( p_0 (t) + p_J (t) \right), \label{eq:EoMredpJ}
\end{equation}
we can also eliminate $\phi_J(t)$ to arrive at two coupled time-delay differential equations for $p_0(t)$ and $p_J(t)$ only. %
\subsubsection{Low impedance TL}
We now proceed to analyze the regime of a low-impedance TL $(\eta < 1)$ in more detail. To receive an analytical solution for the equations of motion Eqs.~\eqref{eq:EoMredpJ1}-\eqref{eq:EoMredpJ}, we rewrite the charge on the coupling capacitance $p_0(t)$ as the corresponding charge for the undamped LC-oscillator (Eq.~(\ref{Eq:P0forZ0eq0})) plus a small perturbation $\delta p_0(t)$: 
\begin{align}
	p_0(t)  = - \frac{C_c}{C_c + C_J} p_J(t) - \delta p_0(t).
	\label{eq:p0Ansatz}
\end{align}
Using this ansatz, Eq.~\eqref{eq:EoMredpJ} becomes
\begin{align}
	\partial_t^2 p_J = - \omega_0^2 p_J + \omega_J^2 \delta p_0,
	\label{Eq:d2tpJ}
\end{align}
where again $\omega_0 = 1/ \sqrt{L_J (C_c + C_J)}$ is the resonance frequency of the qubit coupled to the TL and $\omega_J = 1/ \sqrt{L_J  C_J}$ is the resonance frequency of the uncoupled qubit. From Eq.~\eqref{eq:EoMredpJ1} we find
\begin{align}
	\delta p_0 (t) &= - \frac{Z_0}{2} C_J \left( \frac{C_c}{C_c + C_J} \right)^2 \partial_t \left(p_J (t) - p_J (t - T) \right)  \nonumber \\
	&- \frac{Z_0}{2} \frac{C_c C_J}{C_c + C_J} \partial_t \left( \delta p_0 (t) - \delta p_0 (t - T) \right),
	\label{eq:deltap0approx1M}
\end{align}
where we will now neglect the second term, assuming that the time dependence of $\delta p_0 (t)$ is not qualitatively faster than the one of $p_J(t)$. This gives an expression for $\delta p_0 (t)$ in terms of $p_J(t)$ and $p_J(t-T)$, which inserted in in Eq.~(\ref{Eq:d2tpJ}) gives 
\begin{align}
	\partial_t^2 p_J (t) &= - \omega_0^2 p_J (t) - \gamma_0 \partial_t \left( p_J (t) - p_J (t - T)  \right),
	\label{eq:pjTD}
\end{align}
where we again find the low-impedance decay rate $\gamma_0$ from Eq.~(\ref{Eq:gammaLowEta}).

%
%
%
Thus we have found an approximate equation of motion which only contains the charge $p_J$ on the Josephson junction. This equation can be solved analytically by using a Laplace transformation. This solution is presented in Appendix A. In section \ref{sysres}, we will see that this is the equation that corresponds to the system-bath approach from quantum optics. However, below we see that there are regimes where the full equations including both $p_J$ and $p_0$ give significantly different decay dynamics. 

\subsubsection{Numerical results}
In the following, we will initialize the oscillator/qubit at time $t=0$ with a finite charge $p_J$ at $t=0$, while putting $p_0(t)=0$ for $t\leq 0$. This models switching on the coupling between the qubit and the TL and $t=0$ by adding $C_c$ in this moment. Quantum mechanically, this initial condition corresponds to a coherent state of the oscillator, rather than a single photon excitation. The transient dynamics of the energy relaxation will however be the same in the weak coupling regime, as we show below in the comparison with the system-reservoir approach.   

\begin{figure}[t]
	\begin{minipage}{0.5\textwidth}
		\begin{overpic}[width=1\textwidth]{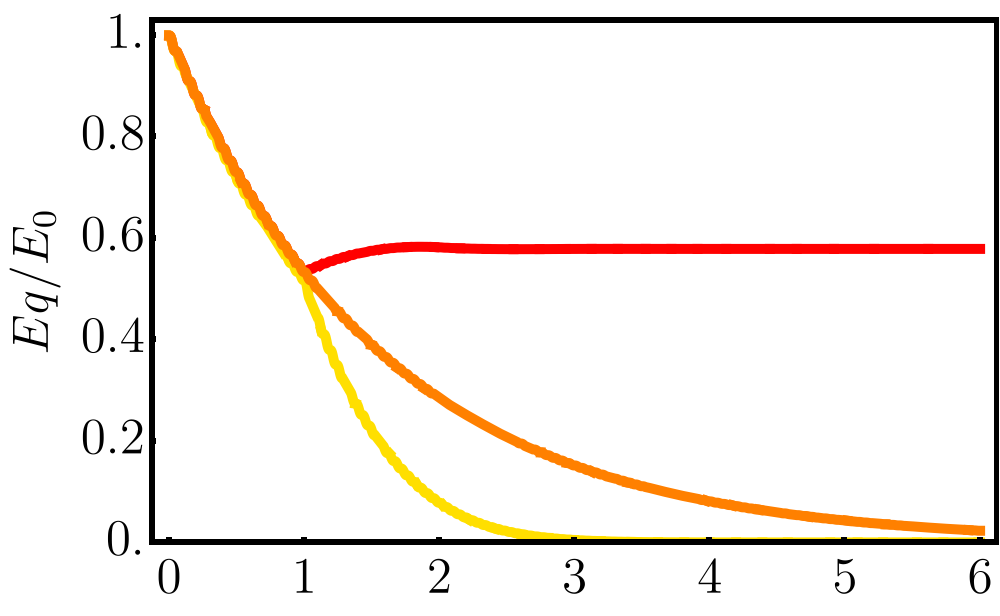}
			\put(3,55){a)}
		\end{overpic}
	\end{minipage}
	\begin{minipage}{0.5\textwidth}
		\begin{overpic}[width=1\textwidth]{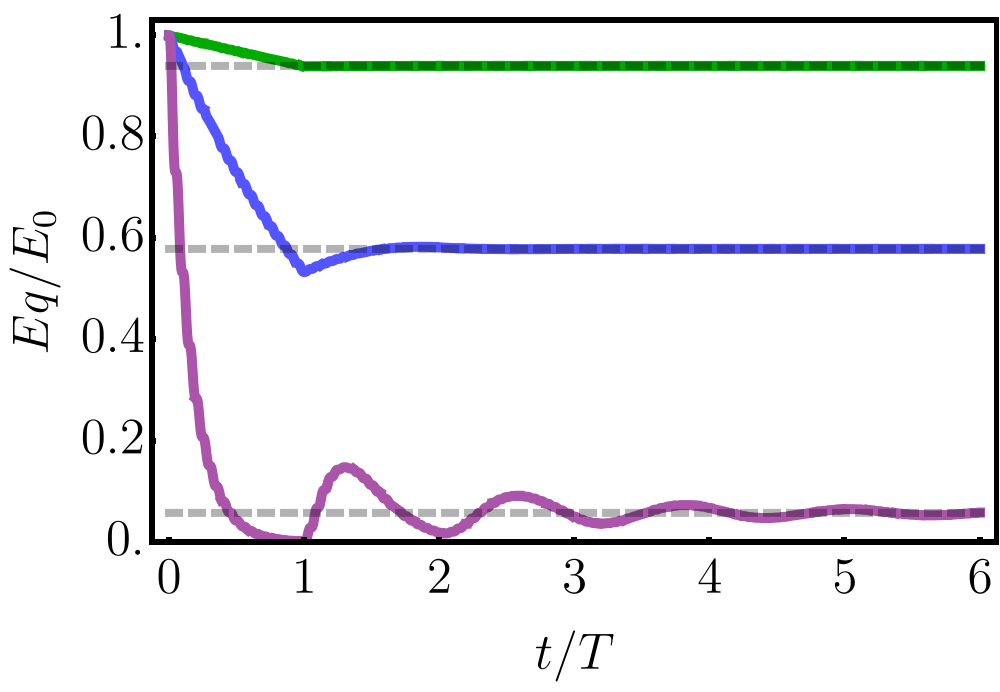}
			\put(3,64){b)}
		\end{overpic}
	\end{minipage}
	\caption{a) The energy of the qubit in front of a mirror for the qubit located at a node (red), at an anti-node (yellow) and the qubit in an open transmission line. If the qubit is located at a node, the energy converges into a dark state. At an anti-node, the decay becomes enhanced by the reflected field from the mirror. In the open transmission line, we see an exponential decay (orange). b) The energy of the qubit located at a node for $\gamma_0 T = 0.01 \cdot 2 \pi $ (green), $\gamma_0 T = 0.1 \cdot 2 \pi $ (blue) and $\gamma_0 T = 1 \cdot 2 \pi $ (purple). In all cases the energy converges into a dark state given by Eq.~(\ref{DarkStateEnergy}) (dashed gray lines), but the transient behaviour is different.}
	\label{fig:EnergyAll}
\end{figure}

We then calculate the energy of the qubit 
\begin{equation}
E_q(t)=\frac{\left(p_J(t)+p_0(t)\right)^2}{2C_J}+\frac{p_0(t)^2}{2C_c} + \frac{\phi_J(t)^2}{2L_J},
\label{Eq:EqforZ0eq0}
\end{equation} 
by solving the equations of motion~\eqref{eq:EoMredpJ1}-\eqref{eq:EoMredpJ} and using $\phi_J(t)=-L_J \partial_t p_J(t)$. 
In Fig.~\ref{fig:EnergyAll} a), we plotted this energy as a function of time for two different positions of the qubit with respect to the mirror. As a reference, we also plot (orange line) the exponential decay found in an open TL. Here, we are in the low-impedance regime where the qubit frequency is given by $\omega_0=1/\sqrt{L_J(C_J+C_c)}$ and the decay rate by Eq.~(\ref{Eq:gammaLowEta}). Including the mirror, we still find exponential decay with the same rate during the first round-trip time period $T$. After this time, we see qualitatively different dynamics depending on the position of the qubit. 

If the qubit is located at a distance where the delay time equals half-integer number of qubit oscillation periods, $T\omega_0=(2n+1)\pi$ for integer $n$, the decay rate increases after time $T$ when the reflected field interacts with the qubit again (yellow line). This occurs when the two terms $p_0(t)$ and $p_0(t-T)$ interfere constructively in Eq.~(\ref{eq:EoMredpJ1}) and correspond to placing the qubit at an anti-node of the electric field at the qubit frequency $\omega_0$.

In this paper, we are however mainly interested in the third case, where the qubit is located at a node of the field (red line), i.e. for $T\omega_0=2n \pi$. In this case, the energy converges into a dark state because the reflected field from the mirror interferes destructively with the outgoing field at any given time.
\subsubsection{Dark-state transients}
The energy remaining in the dark state $E_{DS}$ is given by (see also Eq.~(31) from Ref.~\cite{GonzCira17})
\begin{align}
\frac{E_{DS}}{E_0} =  \frac{1}{(1+ \frac{T }{2}\gamma_0)^2},
\label{DarkStateEnergy}
\end{align}
which we found by calculating the steady-state solution of $p_J$ from Eq.~(\ref{eq:pjTD}), using the Laplace transform solution given in Appendix A. 
We normalized the energy by its initial value $E_0 = E_q(t=0)$ and the factor $\gamma_0 = \frac{Z_0 \omega_0^2}{2}\frac{C_c^2}{C_c + C_J}$ is again the low impedance coupling strength between the qubit and the TL. This energy is shown by dashed lines in Fig.~\ref{fig:EnergyAll}.

In Fig.~\ref{fig:EnergyAll} b) we plotted the energy of the qubit for different values of $\gamma_0 T$. For $\gamma_0 T \ll 1$, the atom decays slowly on the delay time-scale. Then not much of the initial energy is lost until the reflected field from the mirror interacts destructively with the field emitted from the atom and the system reaches the dark state quickly. For $\gamma_0 T \approx 1$, the qubit couples strongly enough to the TL so that it has time to decay significantly before the reflected field interacts with it again. It takes several roundtrips until the emitted and reflected field cancel each other completely and the system reaches a dark state.

\subsubsection{Short outlook towards larger impedance TL}
\begin{figure}[t]
	\begin{minipage}{0.5\textwidth}
		\begin{overpic}[width=1\textwidth]{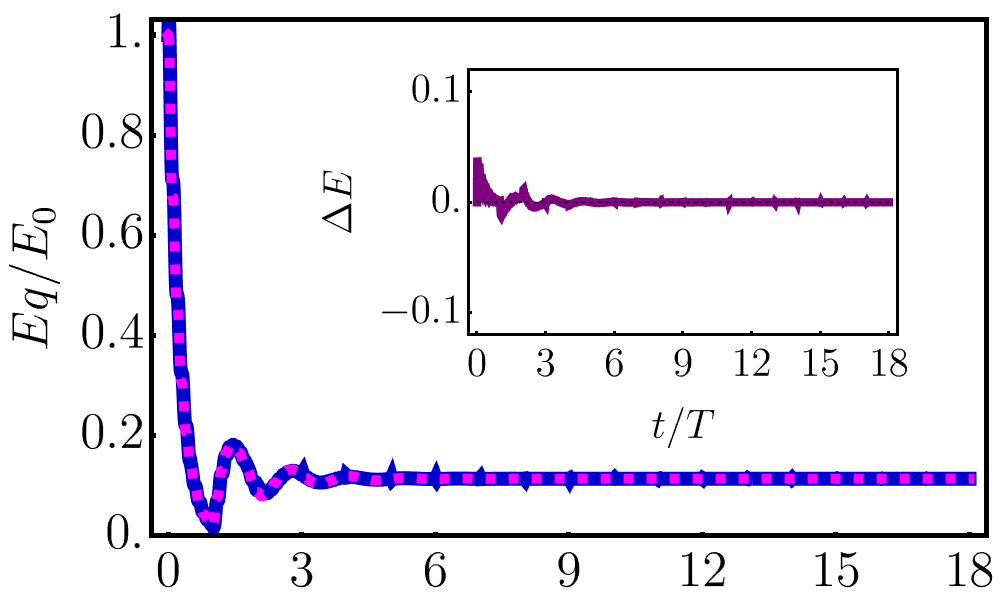}
		\put(3,55){a)}
		\end{overpic}
	\end{minipage}
	\begin{minipage}{0.5\textwidth}
		\begin{overpic}[width=1\textwidth]{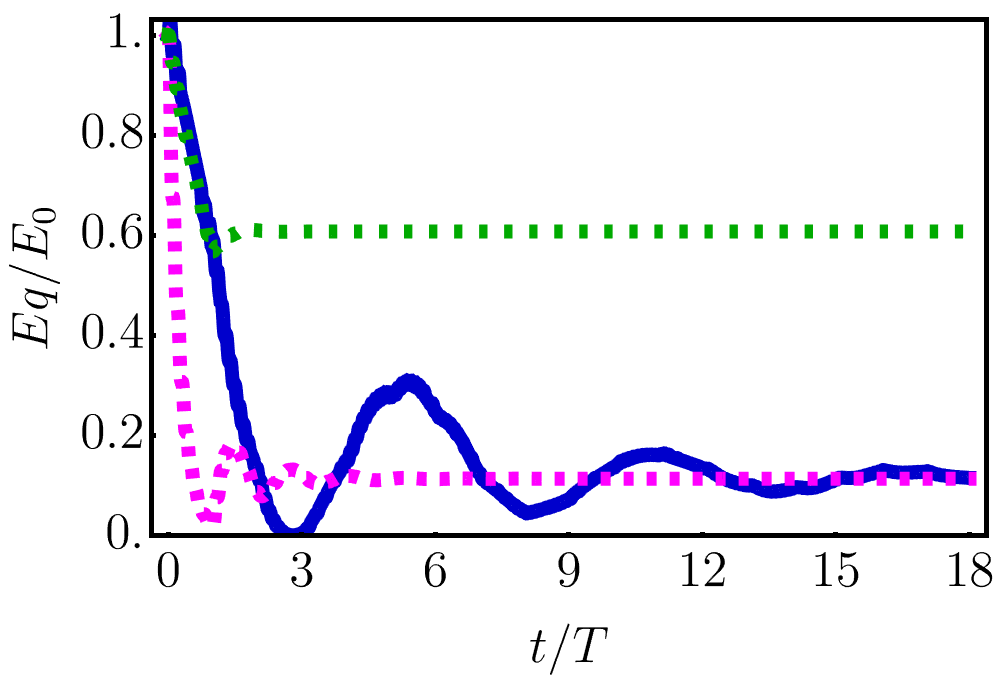}
		\put(3,64){b)}
		\end{overpic}
	\end{minipage}
	\caption{Energy of the transmon qubit as a function of time. In both figures, the value of the coupling is the same $\gamma / \omega_0 = 0.125 $, but in the figure on the top, the impedance is small $Z_0/Z_J =1/\sqrt{2}$ and $\frac{C_c}{C_c + C_J} = 0.5$ and on the bottom figure the impedance is high $Z_0/ Z_J = 100$ and $\frac{C_c}{C_c + C_J} = 0.05$. The pink dashed curve shows the solution of the approximation and the blue curve shows the solution of the full equations. The inset of a) shows the difference of the energy of both cases. We see that for small impedance our approximation works very well, whereas for high impedance the dynamics of the system changes and we can not use the approximation any more. Note, that the dark state energy has the same value either way. The green dashed line in b) shows the energy calculated with the approximation using the value of the coupling strength derived for the high $Z_0$ case.}
	\label{fig:CompareZ0}
\end{figure}

Lately there has been growing interest in high-impedance transmission lines, which can be realized using Josephson junctions or high-kinetic inductance materials in the center conductor \cite{Weissl2015,Krupko2018,Masluk2012}. To study the effect of increasing $Z_0$, we compare the solution of the approximation Eq.~\eqref{eq:pjTD} to the solution of the full equations~\eqref{eq:EoMredpJ1}-\eqref{eq:EoMredpJ}. Figure~\ref{fig:CompareZ0} shows both solutions for two cases with the same value for the low $Z_0$ expression for the coupling $\gamma_0$. In subpanel a) the TL impedance is small, $Z_0/Z_J \leq 1$, and in subpanel b) the TL impedance is high, $Z_0/Z_J \gg 1$, where $\gamma_0$ is kept constant by reducing $C_c$ in the high $Z_0$ case. We see that for small $Z_0$, the approximation describes the behaviour of the energy relaxation very well. For high $Z_0$, we see a big deviation of the full model to the approximation. The source of the deviation becomes clear if we look at Eq.~\eqref{eq:deltap0approx1M}. In the approximation, we neglect the second term. But if we keep $\gamma/\omega_0 \propto Z_0 C_c^2$ constant and increase $Z_0$, which means we decrease $C_c$, it implies that the first term of Eq.~\eqref{eq:deltap0approx1M} becomes small compared to the second term and the second term can therefore not be neglected. One clear difference that is visible in Fig.~\ref{fig:CompareZ0} b) is that the approximation initially decays much faster, which can be understood from the fact that $\gamma_0$ is a low $Z_0$ approximation to the full expression of the open TL decay rate in Eq.~(\ref{FullOpenTLDecayRate}), inadequate for the current parameter regime $\eta > 1$. As a comparison, we therefore plot the solution for the approximate equation of motion Eq.~(\ref{eq:pjTD}), replacing $\gamma_0$ with the full expression for $\gamma$ from Eq.~(\ref{FullOpenTLDecayRate}). This solution, see dashed green curve in Fig.~\ref{fig:CompareZ0} b), captures the initial decay perfectly, but then quickly saturates into a dark state, with much higher energy than the full solution.
The value of the dark-state energy is instead correctly captured by the low impedance approximation in Eq.~(\ref{DarkStateEnergy}), which we also verified analytically in Appendix A, using Laplace transformation of the full equations of motion. 
In the transient dynamics we see oscillations on a new time-scale, arising from energy going back and forth between the qubit and the field between the qubit and the mirror. The detailed analysis of this phenomenon is outside the scope of the current manuscript, but we conclude that dynamics in this regime cannot be captured by the approximate equations of motion Eq.~(\ref{eq:pjTD}), because we need to retain the charge $p_0(t)$ on the coupling capacitance as an independent variable. 
\section{Analogy with the system-reservoir approach}
\label{sysres}
In this section, we start from the circuit-QED Hamiltonian of the system in the continuum limit and connect to a quantum optical system-reservoir approach, where both the transmon qubit and the TL degrees of freedom are quantized. In this model, one degree of freedom of the qubit is directly coupled to the field amplitude in one point and it has been used frequently in literature \cite{Dorner2002, Guo2017}. We find a direct connection between this model and the above equations of motion in the low impedance TL regime. 

  \subsection{Hamiltonian}
\begin{figure}
\includegraphics[width=\columnwidth]{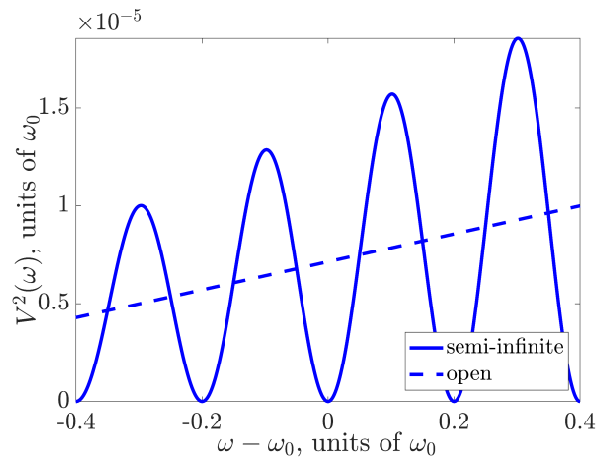}
\caption{Frequency-dependent coupling strength of the transmon versus frequency detuning. Here we chose $\gamma/\omega_0 = 0.05$ and the mirror position is $L = 5\pi v/\omega_0$. The solid line corresponds to the semi-infinite TL with a mirror, while the dashed line corresponds to the open TL case.}
\label{DOS_TL}
\end{figure}

 The Hamiltonian \eqref{Hphip} written in the continuous limit has the form (see Appendix \ref{quantTLt}):
\begin{align}
\label{H_full}
H &= \int{\rm d}x\Big(\frac{p(x)^2}{2C_0} + \frac{1}{2L_0}\Big(\frac{\partial \phi(x)}{\partial x}\Big)^2\Big)\nonumber\\
&+ \frac{p_J^2}{2C_J} + {\cal V}(\phi_J) - \frac{C_c + C_J}{2C_cC_J}p_0^2 + \frac{p(0)}{C_0}p_0.
\end{align}
It should be noted that this Hamiltonian corresponds to the full equations of motion that were solved in previous sections. Because it contains terms in $p_0$, one cannot draw a straighforward analogy with a system-reservoir approach at this stage. To do so, we consider the characteristic impedance $Z_0$ of the TL, and write the relation between the voltages:
\begin{align}
\dot{\phi}_0 = \left|\frac{iZ_0C_c\omega/2}{1 + iZ_0C_c\omega/2}\right|\dot{\phi}_J.
\end{align}
We see from this relation that for $Z_0C_c\omega/2\ll 1$, i.e. for low impedance TLs, the voltage at the 0 node is very small and can be neglected in Eq. \eqref{eq:EoMphi0a}, leading to:
\begin{align}
p_0 \approx -\frac{C_c}{C_c + C_J}p_J.
\end{align}
As a consequence, the charge $p_0$ reveals the TL-transmon coupling term and a frequency shift for the transmon qubit in the Hamiltonian:
\begin{align}
\label{H_cont_x}
H &= \int{\rm d}x\Big(\frac{p(x)^2}{2C_0} + \frac{1}{2L_0}\Big(\frac{\partial \phi(x)}{\partial x}\Big)^2\Big)\nonumber\\
&+ \frac{p_J^2}{2(C_c+C_J)} + {\cal V}(\phi_J) - \frac{C_c}{C_c + C_J}\frac{p(0)}{C_0}p_J.
\end{align}
The TL and transmon degrees of freedom can be quantized as a single harmonic oscillator (since we linearized the transmon qubit) coupled to a reservoir of harmonic oscillators. A rigorous quantization procedure is presented in Appendix \ref{quantTLt} and leads to the rotating wave approximation Hamiltonian being described in terms of creation and annihilation operators:
\begin{align}
\label{H_qRWA}
\op{H} &= \hbar\omega_0\op{a}_J^\dagger\op{a}_J + \intzp{\rm d}\omega\,\hbar\omega\,\op{a}^\dagger(\omega) \op{a}(\omega)\nonumber\\
&+ \intzp{\rm d}\omega\,\hbar V(\omega)\big(\op{a}_J\op{a}^\dagger(\omega) + \op{a}_J^\dagger\op{a}(\omega) \big),
\end{align}
where $\op{a}_J$ annihilates one transmon qubit excitation and $\op{a}(\omega)$ annihilates a sine mode of the TL at frequency $\omega$. The third term on the right-hand side corresponds to the coupling of the transmon with the TL, where:
\begin{align}
V(\omega) = \sqrt{\frac{\gamma}{2\pi}\frac{\omega}{\omega_0}}\sin\frac{\omega L}{v},
\end{align}
where $\gamma$ is the open TL transmon decay rate. Studying the frequency-dependent coupling leads to the Purcell picture, whereby an atom's decay rate is modified by the mode structure of its environment \cite{Purcell}. In Fig.~\ref{DOS_TL}, we show the squared coupling strength, which is proportional to the Purcell factor. We compare it to the open TL coupling strength, which in 1D is just a straight line. Noticeably, the shorted TL case leads to an oscillating coupling depending on the position of the atom with respect to the mirror and the transition frequency $\omega_0$, yielding the transmon decaying as $\e^{-2\gamma t}$ when it is placed at an antinode, while virtually not decaying at all when placed at a node.
  \subsection{Single-excitation basis state evolution}
We study the dynamics of Hamiltonian \eqref{H_qRWA}, assuming that the initial state contains one excitation. Therefore we write the wavefunction in the interaction picture:
\begin{align}
|\psi(t)\ket = c_J(t)\e^{i\omega_0t}|1_J,\vac_{TL}\ket + \intzp\dd\omega\,c_\omega(t)\e^{i\omega t}|\vac_J,1_\omega\ket,
\end{align}
where we introduced the state notations:
\begin{subequations}
\begin{align}
|1_J\ket &= \op{a}_J^\+|\vac_J\ket,\\
|1_\omega\ket &= \op{a}^\+(\omega)|\vac_{TL}\ket,
\end{align}
\end{subequations}
where $|\vac_J\ket,|\vac_{TL}\ket$ are the vacuum states of the transmon and the TL, respectively. Writing down the time-dependent Schr\"odinger equation, we can write the system of equations governing the evolution of the wavefunction coefficients:
\begin{subequations}
\begin{align}
&\dot{c}_J = i\intzp\dd\omega\,V(\omega)\e^{i(\omega - \omega_0)t}c_\omega(t) \\
&\dot{c}_\omega = i V(\omega)\e^{-i(\omega - \omega_0)t}c_J(t).
\end{align}
\end{subequations}
Integrating formally the equations on $c_\omega(t)$, replacing in the equation on $c_J(t)$ and choosing the initial conditions to be $c_J(0) = 1, c_\omega(0) =0$, we have now:
\begin{align}
\label{c_J}
\dot{c}_J = -\int_0^t\dd\tau\,c_J(\tau)\intzp\dd\omega\,\e^{i(\omega-\omega_0)(t-\tau)}V^2(\omega).
\end{align}
To solve this equation, one needs to evaluate the integral over frequencies. Changing the variable to $\Delta = \omega - \omega_0$ and considering that the decay is much smaller than the transition frequency $\gamma\ll\omega_0$, one can extend the lower bound of the integral to $-\infty$ and we get:
\begin{align}
\label{c_J2}
\dot{c}_J = -\frac{\gamma}{\pi\omega_0}\int_0^t\dd\tau\,c_J(\tau)\intmp\dd\Delta\,\e^{i\Delta(t-\tau)}(\Delta + \omega_0)\sin^2\frac{\Delta L}{v},
\end{align}
where we used the fact that the transmon is at a node so $\sin(\Delta L/v + n\pi) = -\sin\Delta L/v$. The right-hand side integral then has the form of a Fourier transform of two terms: one is $\Delta$ times a squared sine, which is an odd function, so only the sine component of $\e^{i\Delta(t-\tau)}$ is non-vanishing. This leads to the integral over a function whose Taylor expansion around $\Delta = 0$ is of the order of ${\cal O}(\Delta^4)$, and since only frequencies around $\omega_0$ will contribute, this term can be considered negligibly small. The remaining term is the Fourier transform of the squared sine, leading to:
\begin{align}
\pi\delta(t-\tau) - \frac{\pi}{2}\delta(t - \tau - T) - \frac{\pi}{2}\delta(t-\tau + T),
\end{align}
where $T= 2L/v$. The equation of motion then becomes simply:
\begin{align}
\dot{c}_J = -\frac{\gamma}{2}\big(c_J(t) - c_J(t - T)\big).
\end{align}
This equation is in the interaction picture, but the Schr\"odinger picture can be obtained by changing the rotating frame: $c_J(t) = \widetilde{c}_J(t)\e^{i\omega_0 t}$:
\begin{align}
\dot{\widetilde{c}}_J = -i\omega_0\widetilde{c}_J(t)-\frac{\gamma}{2}\big(\widetilde{c}_J(t) - \widetilde{c}_J(t - T)\e^{-i\omega_0 T}\big).
\end{align}
Again we can consider the atom being at a node so that $\omega_0T = 2 n \pi$, and the phase factor in the last term is then just 1. This result is consistent with the derivation shown in refs. \cite{TufaKim13,Guo2017} and leads to the same dynamics. 

However it is crucial to note that the behaviour of the qubit energy in the case of high impedance \emph{cannot} be modeled with this approach. Our semi-classical analysis revealed non-Markovian oscillations for the energy with $Z_0/Z_J\gg 1$, as shown in Fig.~\ref{fig:CompareZ0}, and those cannot be captured by the weak coupling and low impedance system-reservoir model derived in this section. To derive a proper quantum approach, one should come back to Hamiltonian \eqref{H_full} and derive the equations of motion for the full system including the charge $p_0$.

\section{Fast-oscillating terms}
\label{sec:Wiggles}
\begin{figure}[t]
	\begin{overpic}[width=1\linewidth]{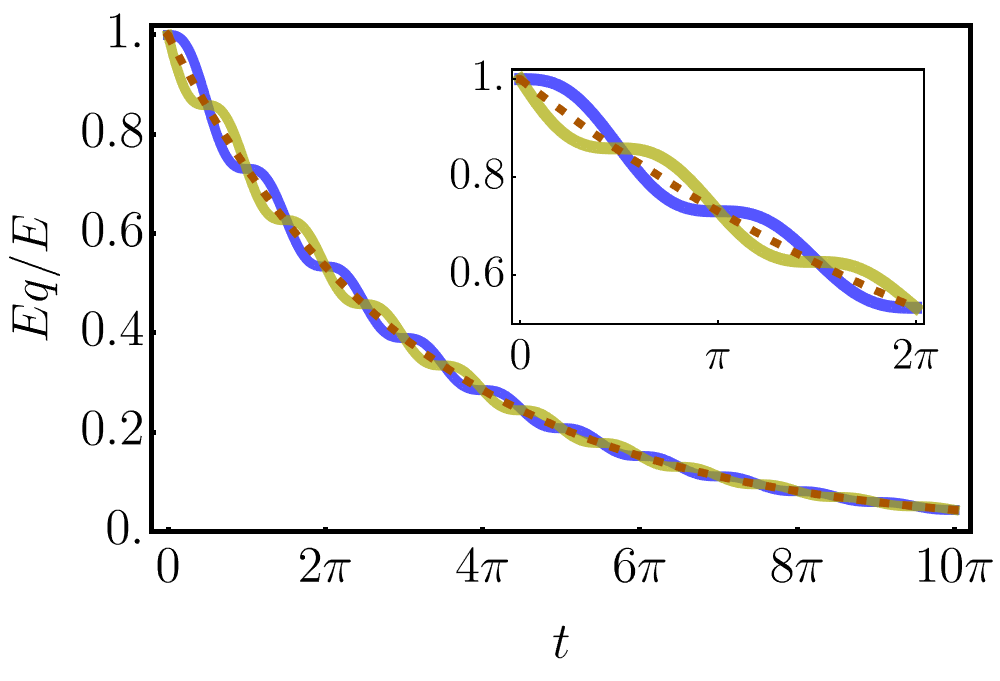}
		\put(49,3){\large $\omega_0$}
	\end{overpic}
	\caption{Energy of an initially excited qubit in an open transmission line. The blue and green line show the decay of the qubit for $\gamma/\omega_0 = 0.1$. In this parameter regime we can observe the fast-oscillating behaviour. The phase of the oscillations depend on the initial conditions, where $p_J(0)$ is finite and $p_{0}(0)=\phi_{J}(0)=0$ for the blue curve and $\phi_{J}(0)$ is finite and $p_J(0) = p_{0}(0)$ for the green curve.  For the red dashed line, the coupling is significantly smaller than the resonance frequency of the qubit $\gamma / \omega_0 = 0.001$ and the fast oscillations can not be seen any more. The inset shows a magnification for the first period.}
	\label{fig:Wiggles}
\end{figure}

Usually, when dealing with emitters coupled to an electromagnetic field, the rotating wave approximation is used and fast rotating terms are neglected. However, in our semi-classical model we are not doing the rotating wave approximation and see effects of the fast rotating terms. To demonstrate the behaviour of these terms, we analytically solve the equations for an atom in an open transmission line. In this case, the time delay term in Eq.~\eqref{eq:pjTD} is not present and the equation can be reduced to
\begin{align}
	\partial_t^2 p_J^{\text{TL}}  &= - \omega_0^2 p_J^{\text{TL}}  - \gamma \partial_t  p_J^{\text{TL}} .
\end{align}
For $\gamma/\omega_0 \ll 1$, the solution of this equation is given by
\begin{align}
	\tilde p_J = \frac{p_J^{\text{TL}}}{P_J(0)} = e^{- \frac{\gamma}{2} t} \cos \left(\omega_0 t\right).
\end{align}
The energy of the qubit can then be written as
\begin{align}
	E/E_0 &= \tilde p_J^2 + \frac{1}{\omega_0^2} \tilde \phi_J^2 \nonumber \\
			  &= e^{- \gamma t} \left[ \cos^2 (\omega_0 t) \left( 1 + \frac{\gamma^2}{4 \omega_0^2}  \right)  \right. \nonumber\\ 
			  & \qquad \left. +  \sin^2(\omega_0 t) + \frac{\gamma }{2\omega_0} \sin (2 \omega_0 t)  \right]\nonumber \\
			  &= e^{- \gamma t} \left[1 + \frac{\gamma }{2\omega_0} \sin (2 \omega_0 t) + \frac{\gamma^2}{4 \omega_0^2}\cos^2 (\omega_0 t)\right],
\end{align}
where we can see that the last two terms oscillate with the frequency $2 \omega_0$, which corresponds to the fast rotating terms. The terms that contain the fast oscillations are proportional to the factor $\gamma/\omega_0$ and $(\gamma/\omega_0)^2$, respectively. For weak coupling $\gamma / \omega_0 \ll 1$, the oscillations are not visible (see dashed red curve in Fig.~\ref{fig:Wiggles}, where $\gamma/\omega_0 = 0.001$). The blue curve in Fig.~\ref{fig:Wiggles} shows the energy of the qubit for $\gamma/\omega_0 = 0.1$, which is significantly larger than for the other case and the fast oscillations are clearly visible. Here, we note that the phase of these fast oscillations depends on the initial state, which in our case is chosen to be a finite $p_J(t=0)$ while $p_0(0)=\phi_J(0)=0$. Choosing instead a finite $\phi_J(t=0)$ shifts the oscillations $\pi/2$, see the green curve in in Fig.~\ref{fig:Wiggles}. A single-photon Fock state has an undetermined phase, so averaging over the initial phase to mimic this quantum initial state would indeed wash out these fast oscillations. However, to fully analyse the effects of these counter-rotating terms in the ultrastrong coupling regime where $\gamma/\omega_0 \sim 1$ is beyond the scope of this manuscript.
\section{Conclusion}
We have investigated the spontaneous emission dynamics of an initially excited superconducting artificial atom of transmon type, capacitively coupled to a semi-infinite transmission, shorted at a distance $L$ from the transmon. Using a circuit quantization procedure, we derived time-delay equations of motion for the charge on the transmon and on the coupling capacitance. Replacing the Josephson junction by its Josephson inductance, we arrived at linear equations of motion. The average charges then obey identical scalar equations of motion, which we then proceed to solve. We found that the energy relaxation depends strongly on the distance between the atom and the mirror, in terms of the wavelength of the emitted radiation. We especially focused on the case where the atom is located at a node of the electro-magnetic field, leading the atom to converge into a dark state with finite energy in the steady state. We found a simple analytical expression for this energy. We then found very different dynamics depending on the characteristic impedance of the transmission line compared to the characteristic impedance of the transmon. For a small transmission line impedance we found an approximate equation of motion for the atom charge only. In this regime, we could also derive the corresponding equations of motion of a single emitter in a quantum optical system-bath approach, previously used in literature. However, in the regime of large characteristic impedance of the transmission line, we found that the charge on the coupling capacitance must be retained as a separated degree of freedom and the mapping to a quantum optical model is not clear. We have thus established a solid connection between the circuit-QED model and the quantum optical master equation approach in the regime of small characteristic impedance of the transmission line. We have also established a framework in which one can perform a detailed analysis of the high-impedance regime. 
\label{conclusion}
\begin{acknowledgments}
    The authors thank Luis Martin Moreno for stimulating discussions and Mikhail Pletyukhov for assistance with the Laplace transformation. We also thank the Swedish Research Council and the Knut and Alice Wallenberg foundation for financial support.
\end{acknowledgments}
\newpage
\appendix
\section{Laplace Transform}
To calculate the energy of the dark state, we want to find the Laplace transform of $p_J$ and $p_0$. Therefore we do the Laplace transformation of the following equations
\begin{align} 
	\frac{C_{c}+C_{J}}{C_{c} C_{J}} p_{0}(t)+\frac{1}{C_{J}} p_{J}(t) &=-\frac{Z_{0}}{2} \partial_{t}\left(p_{0}(t)-p_{0}(t-T)\right), \\ 
									 \partial_{t}^{2} p_{J}(t)&=-\omega_{J}^{2}\left(p_{0}(t)+p_{J}(t)\right) 
\end{align}
The Laplace transform of these equations is given by
\begin{widetext}
\begin{align}
			\frac{C_{c}+C_{J}}{C_{c} C_{J}} \tilde p_{0}(s) +\frac{1}{C_{J}} \tilde p_{J}(s) &=-\frac{Z_{0}}{2} \left( s \tilde p_{0}(s) \left(1 - e^{-s T}\right)  + p_0(0) - p_0(-T) \right), \\ 
					 s^2 \tilde p_{J}(s) - s p_J (0) - p_J' (0) &=-\omega_{J}^{2}\left( \tilde p_{0}(s)+ \tilde p_{J}(s)\right) 
\end{align}
and we find
\begin{align}
	\tilde p_0 (s) = -\frac{2 C_c  L_J p_J(0) s e^{sT} } {-C_c s\left(1+C_J L_J s^{2} \right) Z_0 + e^{s T} \left(2+2C_J L_J s^2+C_c s (Z_0+L_J s (2+C_J s Z_0))\right)}, \\
	\tilde p_J (s) = -\frac{ p_J (0) s L_J \left( -C_c C_J s Z_0 + e^{s T} \left( 2 (C_c + C_J) + C_c C_J Z_0 s \right)  \right) } {-C_c s\left(1+C_J L_J s^{2} \right) Z_0 + e^{s T} \left(2+2C_J L_J s^2+C_c s (Z_0+L_J s (2+C_J s Z_0))\right)},
\end{align}
\end{widetext}
where we assumed that $p_J' (0) = p_0(0) = p_0(-T)=0$. To calculate the energy of the dark state, we use the dark state condition $\omega_0 T = 2 \pi n$ and calculate the steady state condition for the Laplace transform, which is given by
\begin{align}
\lim _{t \rightarrow \infty} f(t)=\lim _{s \rightarrow 0} s \mathcal{F}(s).
\end{align}
We find
\begin{align}
\lim _{t \rightarrow \infty} p_{J}(t) &= \lim _{s \rightarrow 0} s \tilde{p}_{J}\left(s+i \omega_{0}\right)+\lim _{s \rightarrow 0} s \tilde{p}_{J}\left(s-i \omega_{0}\right) \\
&= p_J (0) \frac{1}{1 + \frac{\gamma_0}{2} T}
\end{align}
and
\begin{align}
	\lim _{t \rightarrow \infty} p_{0}(t) &= \lim _{s \rightarrow 0} s \tilde{p}_{0}\left(s+i \omega_{0}\right)+\lim _{s \rightarrow 0} s \tilde{p}_{0}\left(s-i \omega_{0}\right) \\
	&= \frac{-C_c}{C_c + C_J} p_J (0) \frac{1}{1 + \frac{\gamma_0}{2} T}.
\end{align}
The energy of the dark state is then given by
\begin{align}
  \frac{E_{\text{DS}}}{E_{0}}=\frac{1}{\left(1+\frac{\gamma_0}{2} T\right)^{2}}.
\end{align}
\subsection{Analytical solution for low $Z_0$}
The low impedance approximate equation of motion for $p_J(t)$ is given by
\begin{align}
\partial_t^2 p_J (t) &= - \omega_0^2 p_J (t) - \gamma_0 \partial_t \left( p_J (t) - p_J (t - T)  \right).
\end{align}
with $\gamma_0 = \frac{Z_0}{2 } \omega_0^2 \frac{C_c^2}{C_c + C_J}$ and $\omega_0 = 1/ \sqrt{L_J (C_c + C_J)}$. \\
The Laplace transform of this is
\begin{multline}
s^2 \tilde{p}_J(s) - s p_J(0) - p_J'(0) = - \omega_0^2 \tilde{p_J} (s) - \gamma_0 s \tilde{p_J} (s) \\ + \gamma_0 p_J (0) + \gamma_0 s e^{-s T} \tilde{p_J} (s) - \gamma_0 p_J(-T).
\end{multline}
So, we obtain
\begin{align}
\tilde{p_J} (s) = \frac{(\gamma_0 + s) p_J (0)}{s^2 + \gamma_0 s ( 1 - e^{-s T }) + \omega_0^2},
\label{eq:LaplacepJ}
\end{align}
where we assumed that $p_J'(0) = 0$ and $p_J(-T) = 0$.
This can be rewritten as
\begin{align}
		\tilde { p } _ { J } ( s ) &= p _ { J } ( 0 ) \frac { ( s + \gamma_0 ) } { l ( s ) - \gamma_0 s e ^ { - s T } } \\
										  &= p _ { J } ( 0 ) \frac { ( s + \gamma_0 ) } { l ( s ) } \sum _ { n = 0 } ^ { \infty } \left[ \frac { \gamma_0 s } { l ( s ) } \right] ^ { n } e ^ { - s n T },
\end{align}
with
\begin{align}
l ( s ) & = s ^ { 2 } + \gamma_0 s + \omega _ { 0 } ^ { 2 } = \left( s - s _ { + } \right) \left( s - s _ { - } \right) \\ 
s _ { \pm } & = - \frac { \gamma_0 } { 2 } \pm \frac { \alpha } { 2 } , \quad \alpha = 2 \sqrt { \left( \frac { \gamma_0 } { 2 } \right) ^ { 2 } - \omega _ { 0 } ^ { 2 } }.
\end{align}
The integral for the inverse Laplace transform reads
\begin{align}
\frac { p _ { J } ( t ) } { p _ { J } ( 0 ) } = \frac { 1 } { 2 \pi i } \sum _ { n = 0 } ^ { \infty } \int _ { - i \infty } ^ { i \infty } \frac { \gamma_0 ^ { n } s ^ { n } ( s + \gamma_0 )e ^ { s ( t - n T ) } \diff s } { \left( s - s _ { + } \right) ^ { n + 1 } \left( s - s _ { - } \right) ^ { n + 1 } } .
\label{eq:pJ}
\end{align}
To solve this, we define
\begin{align}
f(s) = \frac {  s ^ { n } ( s + \gamma_0 ) } { \left( s - s _ { + } \right) ^ { n + 1 } \left( s - s _ { - } \right) ^ { n + 1 } } e ^ { s ( t - n T ) }
\end{align}
and use the residue theorem
\begin{align}
\oint_K f(z) dz = 2 \pi i \sum_{k=0}^{n} \left. \mathop{Res} f(z) \right|_{z=z_k},
\end{align}
where $z_k$ are the poles of $f(z)$ and $\left. \mathop{Res} f(z) \right|_{z=z_k}$ can be written as
\begin{align}
\left. \mathop{Res} f(z) \right|_{z=z_0} = \lim_{z \to z_0} \frac{1}{(m-1)!} \frac{\diff^{m-1}}{\diff z^{m-1}} \left[f(z) (z-z_0)^m\right].
\end{align}
The poles of $f(s)$ are $s_+$ and $s_-$
\begin{align}
\left. \mathop{Res} f(s) \right|_{s=s^+} &= \frac{1}{n!} \left[\frac { d ^ { n } } { d s ^ { n } }  \left( \frac { s ^ { n } ( s + \gamma_0 ) e ^ { s ( t - n T ) } } { \left( s - s _ { - } \right) ^ { n + 1 } } \right) \right]_ { s = s _ { + } } \\
\left. \mathop{Res} f(s) \right|_{s=s^-} &= \frac{1}{n!} \left[\frac { d ^ { n } } { d s ^ { n } }  \left( \frac { s ^ { n } ( s + \gamma_0 ) e ^ { s ( t - n T ) } } { \left( s - s_ { + } \right) ^ { n + 1 } } \right) \right]_ { s = s _ { - } }.
\end{align}
These we can rewrite by shifting $s$ to $s \rightarrow s + s^+$ and $s \rightarrow s + s^-$
\begin{widetext}
\begin{align}
\left. \mathop{Res} f(s) \right|_{s=s^+} &= \frac{1}{n!} e ^ { s_+ ( t - n T ) }   \left[ \frac { d ^ { n } } { d s ^ { n } }\left(\frac { \left( s + s _ { + } \right) ^ { n } \left( s + s _ { + } + \gamma_0 \right) e ^ { s ( t - n T ) } } { \left( s + s _ { + } - s _ { - } \right) ^ { n + 1 } } \right) \right]_ { s = 0 } \\
\left. \mathop{Res} f(s) \right|_{s=s^-} &= \frac{1}{n!} e ^ { s_- ( t - n T ) } \left[ \frac { d ^ { n } } { d s ^ { n } }  \left(\frac { \left( s + s _ { - } \right) ^ { n } \left( s + s _ { - } + \gamma_0 \right) e ^ { s ( t - n T ) } } { \left( s + s _ { - } - s _ { + } \right) ^ { n + 1 } } \right) \right]_ { s = 0 }.
\end{align}
We set this into Eq.~\eqref{eq:pJ} and obtain the solution of the inverse Laplace transform
\begin{multline}
\frac { p _ { J } ( t ) } { p _ { J } ( 0 ) } = \sum_{ n = 0 }^{\infty} \Theta \left(t - n T\right)\frac{\gamma_0^n}{n!} \bigg\{  e ^ { s_+ ( t - n T ) }
\left[ \frac { d ^ { n } } { d s ^ { n } }  \left(\frac { \left( s + s _ { + } \right) ^ { n } \left( s + s _ { + } + \gamma_0 \right) e ^ { s ( t - n T ) } } { \left( s + s _ { + } - s _ { - } \right) ^ { n + 1 } } \right) \right]_ { s = 0 }\\ + e ^ { s_- ( t - n T ) }\left[ \frac { d ^ { n } } { d s ^ { n } }  \left(\frac { \left( s + s _ { - } \right) ^ { n } \left( s + s _ { - } + \gamma_0 \right) e ^ { s ( t - n T ) } } { \left( s + s _ { - } - s _ { + } \right) ^ { n + 1 } } \right) \right]_ { s = 0 } \bigg\}.
\end{multline}
\end{widetext}
\subsection{Steady state}
The steady state solution of a function $f(t)$ is given by
\begin{align}
 \lim_{t \to \infty} f(t) = \lim_{s \to 0} s \mathcal{F} (s),
\end{align}
where $\mathcal{F}(s) = \int_{0}^{\infty} e^{-s t} f(t) \diff t$ is the Laplace transform of $f(t)$.
To calculate an expression for the energy of the dark state, we calculate the steady state solution of $p_J (t)$,
\begin{align}
	\lim_{t \to \infty} p_J(t) = \lim_{s \to 0} s  \tilde p_J (s+ i \omega_0) + \lim_{s \to 0} s  \tilde p_J (s- i \omega_0).
\end{align}
Using the Laplace transform~\eqref{eq:LaplacepJ} this becomes
\begin{widetext}
\begin{align}
	\lim_{t \to \infty} \frac{p_J(t)}{p_J(0)} &= \lim_{s \to 0} s \frac{\gamma_0 + (s+ i \omega_0)}{(s+ i \omega_0)^2 + \gamma_0 (s+ i \omega_0) ( 1 - e^{-(s+ 														i \omega_0) T }) + \omega_0^2} \\
																&+ \lim_{s \to 0} s \frac{\gamma_0 + (s- i \omega_0)}{(s- i \omega_0)^2 + \gamma_0 (s- i \omega_0) ( 1 - e^{-(s- i \omega_0) T }) + \omega_0^2} \\
																&= \lim_{s \to 0} s \frac{\gamma_0 + (s+ i \omega_0)}{(s+ i \omega_0)^2 + \gamma_0 (s+ i \omega_0) ( 1 - e^{-s T + i 2 \pi n}) + \omega_0^2} \\
																&+ \lim_{s \to 0} s \frac{\gamma_0 + (s- i \omega_0)}{(s- i \omega_0)^2 + \gamma_0 (s- i \omega_0) ( 1 - e^{-s T + i 2 \pi n }) + \omega_0^2} \\
																&= - i \frac{\gamma_0 + i \omega_0}{2 \omega_0 + T \gamma_0 \omega_0} + i \frac{\gamma_0 - i \omega_0}{2 \omega_0 + T \gamma_0 \omega_0} \\
																 &= \frac{1}{1 + \frac{\gamma_0}{2} T},
\end{align}
\end{widetext}
where we used the condition for the dark state $ \omega_0 T = 2 \pi n$.
The energy of the qubit in the steady state is given by
\begin{align}
	\frac{E_J}{E_0} = \frac{1}{\left( 1 + \frac{\gamma_0}{2} T \right)^2}.
\end{align}
\section{Quantization of the TL-transmon system}
\label{quantTLt}
Considering the general solutions for the flux $\phi(x,t)$ and the charge density $p(x,t)$, we now derive the Hamiltonian \eqref{H_cont_x} with the quantized modes. The general solutions of the TL modes when the line is grounded at $x=L$ are:
\begin{subequations}
\begin{align}
\phi(x,t) &= \sqrt{\frac{2}{\pi}}\int_{0}^{+\infty}\frac{{\rm d}\omega}{v}\,\phi(\omega,t)\sin \frac{\omega}{v}(x - L),\\
p(x,t) &= \sqrt{\frac{2}{\pi}}\int_{0}^{+\infty}\frac{{\rm d}\omega}{v}\,p(\omega,t)\sin \frac{\omega}{v}(x - L),
\end{align}
\end{subequations}
where $\omega = |k|v$ and $\phi(\omega,t)$, $p(\omega,t)$ are real coefficients on the sine modes. The latter are linked with the Fourier transforms of the general solutions:
\begin{align}
f(\omega,t) = i\widetilde{f}(k,t){\rm e}^{-ikL},
\end{align}
where $f=\phi, p$ and $\widetilde{f}(k,t) = {\cal F}_x[f](k)$ are the Fourier transforms. We write the Hamiltonian \eqref{H_cont_x} with the zero boundary condition at $x=L$. Also, the time dependence of the Hamiltonian due to kinetic and potential term is implicit, and no external time-dependent potential is considered. Therefore, one can set $t=0$ in the expression of the Hamiltonian, and this yields the Schr\"odinger picture. The TL part of the Hamiltonian is then:
\begin{align}
H_{TL} = \int_{-\infty}^L{\rm d} x \Big(\frac{p^2(x,0)}{2C_0} + \frac{1}{2L_0}\Big(\frac{\partial\phi(x,0)}{\partial x}\Big)^2\Big).
\end{align}
The expressions of $\phi(x,t)$ and $p(x,t)$ are now replaced by the general solutions. This brings up terms in $\sin k(x-L)\sin k'(x-L)$ and $\cos k(x-L)\cos k'(x-L)$ which reduce to dirac deltas with the integration over $x$, and we get:
\begin{align}
H_{TL} = \frac{1}{2}\int_0^{+\infty}{\rm d}\omega\Big(Z_0p^2(\omega,0)+ \frac{k^2}{Z_0}\phi^2(\omega,0)\Big),
\end{align}
where $\phi,p(\omega) \equiv\phi,p(\omega,0)$. The canonical variables can now be decomposed into annihilation and creation operators:
\begin{align}
\op{\phi}(\omega) &= \sqrt{\frac{\hbar Z_0v^2}{2\omega}}\big(\op{a}(\omega) + \op{a}^\dagger(\omega)\big)\\
\op{p}(\omega) &= -i\sqrt{\frac{\hbar\omega}{2Z_0}}\big(\op{a}(\omega) - \op{a}^\dagger(\omega)\big),
\end{align}
where here $\op{a}(\omega)$ must have a dimension $\omega^{-1/2}$. The latter must satisfy the commutation relations $[\op{a}(\omega),\op{a}^\dagger(\omega')] = \delta(\omega - \omega')$. We also need the expression of the charge density at $x=0$ to determine the coupling term in \eqref{H_cont_x}:
\begin{align}
\op{p}(0) = i\int_0^{+\infty}{\rm d}\omega\sqrt{\frac{\hbar\omega}{\pi Z_0}}\big(\op{a}(\omega)-\op{a}^\dagger(\omega)\big)\sin\frac{\omega L}{v}.
\end{align}
Finally, the quantization of the transmon qubit is done using:
\begin{align}
\op{\phi}_J &= \sqrt{{\textstyle\frac{\hbar}{2(C_c +C_J)\omega_0}}}\big(\op{a}_J + \op{a}_J^\dagger\big),\\
\op{p}_J &= -i\sqrt{\frac{\hbar}{2L_J\omega_0}}\big(\op{a}_J - \op{a}_J^\dagger\big),
\end{align}
where $\omega_0 = (L_J(C_c+C_J))^{-1/2}$ is the renormalized qubit frequency. The Hamiltonian then has the form:
\begin{align}
\label{H_q}
\op{H} &= \hbar\omega_0\op{a}_J^\dagger\op{a}_J + \int_0^{+\infty}{\rm d}\omega\,\hbar\omega\,\op{a}^\dagger(\omega) \op{a}(\omega)\nonumber\\
&- \int_0^{+\infty}{\rm d}\omega\,\hbar V(\omega)\big(\op{a}_J - \op{a}_J^\dagger\big)\big(\op{a}(\omega) - \op{a}^\dagger(\omega) \big),
\end{align}
where the frequency-dependent coupling is:
\begin{align}
V(\omega) = \frac{C_c}{C_c + C_J}\sqrt{\frac{Z_0}{4\pi L_J}}\sqrt{\frac{\omega}{\omega_0}}\sin\frac{\omega L}{v}.
\end{align}
The Hamiltonian \eqref{H_q} can be written in the rotating wave approximation:
\begin{align}
\op{H}_\text{RWA} &= \hbar\omega_0\op{a}_J^\dagger\op{a}_J + \intzp{\rm d}\omega\,\hbar\omega\,\op{a}^\dagger(\omega) \op{a}(\omega)\nonumber\\
&+ \intzp{\rm d}\omega\,\hbar V(\omega)\big(\op{a}_J\op{a}^\dagger(\omega) + \op{a}_J^\dagger\op{a}(\omega) \big).
\end{align}

\bibliographystyle{apsrev4-1}
\bibliography{PaperDarkState.bbl}

\begin{thebibliography}{42}%
\makeatletter
\providecommand \@ifxundefined [1]{%
 \@ifx{#1\undefined}
}%
\providecommand \@ifnum [1]{%
 \ifnum #1\expandafter \@firstoftwo
 \else \expandafter \@secondoftwo
 \fi
}%
\providecommand \@ifx [1]{%
 \ifx #1\expandafter \@firstoftwo
 \else \expandafter \@secondoftwo
 \fi
}%
\providecommand \natexlab [1]{#1}%
\providecommand \enquote  [1]{``#1''}%
\providecommand \bibnamefont  [1]{#1}%
\providecommand \bibfnamefont [1]{#1}%
\providecommand \citenamefont [1]{#1}%
\providecommand \href@noop [0]{\@secondoftwo}%
\providecommand \href [0]{\begingroup \@sanitize@url \@href}%
\providecommand \@href[1]{\@@startlink{#1}\@@href}%
\providecommand \@@href[1]{\endgroup#1\@@endlink}%
\providecommand \@sanitize@url [0]{\catcode `\\12\catcode `\$12\catcode
  `\&12\catcode `\#12\catcode `\^12\catcode `\_12\catcode `\%12\relax}%
\providecommand \@@startlink[1]{}%
\providecommand \@@endlink[0]{}%
\providecommand \url  [0]{\begingroup\@sanitize@url \@url }%
\providecommand \@url [1]{\endgroup\@href {#1}{\urlprefix }}%
\providecommand \urlprefix  [0]{URL }%
\providecommand \Eprint [0]{\href }%
\providecommand \doibase [0]{http://dx.doi.org/}%
\providecommand \selectlanguage [0]{\@gobble}%
\providecommand \bibinfo  [0]{\@secondoftwo}%
\providecommand \bibfield  [0]{\@secondoftwo}%
\providecommand \translation [1]{[#1]}%
\providecommand \BibitemOpen [0]{}%
\providecommand \bibitemStop [0]{}%
\providecommand \bibitemNoStop [0]{.\EOS\space}%
\providecommand \EOS [0]{\spacefactor3000\relax}%
\providecommand \BibitemShut  [1]{\csname bibitem#1\endcsname}%
\let\auto@bib@innerbib\@empty
\bibitem [{\citenamefont {Roy}\ \emph {et~al.}(2017)\citenamefont {Roy},
  \citenamefont {Wilson},\ and\ \citenamefont {Firstenberg}}]{Roy2017}%
  \BibitemOpen
  \bibfield  {author} {\bibinfo {author} {\bibfnamefont {D.}~\bibnamefont
  {Roy}}, \bibinfo {author} {\bibfnamefont {C.~M.}\ \bibnamefont {Wilson}}, \
  and\ \bibinfo {author} {\bibfnamefont {O.}~\bibnamefont {Firstenberg}},\
  }\href {\doibase 10.1103/RevModPhys.89.021001} {\bibfield  {journal}
  {\bibinfo  {journal} {Rev. Mod. Phys.}\ }\textbf {\bibinfo {volume} {89}},\
  \bibinfo {pages} {021001} (\bibinfo {year} {2017})}\BibitemShut {NoStop}%
\bibitem [{\citenamefont {Gu}\ \emph {et~al.}(2017)\citenamefont {Gu},
  \citenamefont {F.~Kockum}, \citenamefont {Miranowicz}, \citenamefont {Liu},\
  and\ \citenamefont {Nori}}]{Gu2017}%
  \BibitemOpen
  \bibfield  {author} {\bibinfo {author} {\bibfnamefont {X.}~\bibnamefont
  {Gu}}, \bibinfo {author} {\bibfnamefont {A.}~\bibnamefont {F.~Kockum}},
  \bibinfo {author} {\bibfnamefont {A.}~\bibnamefont {Miranowicz}}, \bibinfo
  {author} {\bibfnamefont {Y.-x.}\ \bibnamefont {Liu}}, \ and\ \bibinfo
  {author} {\bibfnamefont {F.}~\bibnamefont {Nori}},\ }\href {\doibase
  10.1016/j.physrep.2017.10.002} {\bibfield  {journal} {\bibinfo  {journal}
  {Physics Reports}\ }\textbf {\bibinfo {volume} {718-719}},\ \bibinfo {pages}
  {1} (\bibinfo {year} {2017})}\BibitemShut {NoStop}%
\bibitem [{\citenamefont {Gonz\'alez-Tudela}\ \emph {et~al.}(2017)\citenamefont
  {Gonz\'alez-Tudela}, \citenamefont {Paulisch}, \citenamefont {Kimble},\ and\
  \citenamefont {Cirac}}]{GonzCira17_2}%
  \BibitemOpen
  \bibfield  {author} {\bibinfo {author} {\bibfnamefont {A.}~\bibnamefont
  {Gonz\'alez-Tudela}}, \bibinfo {author} {\bibfnamefont {V.}~\bibnamefont
  {Paulisch}}, \bibinfo {author} {\bibfnamefont {H.~J.}\ \bibnamefont
  {Kimble}}, \ and\ \bibinfo {author} {\bibfnamefont {J.~I.}\ \bibnamefont
  {Cirac}},\ }\href {\doibase 10.1103/PhysRevLett.118.213601} {\bibfield
  {journal} {\bibinfo  {journal} {Phys. Rev. Lett.}\ }\textbf {\bibinfo
  {volume} {118}},\ \bibinfo {pages} {213601} (\bibinfo {year}
  {2017})}\BibitemShut {NoStop}%
\bibitem [{\citenamefont {Paulisch}\ \emph {et~al.}(2018)\citenamefont
  {Paulisch}, \citenamefont {Kimble}, \citenamefont {Cirac},\ and\
  \citenamefont {Gonz\'alez-Tudela}}]{PaulGonz18}%
  \BibitemOpen
  \bibfield  {author} {\bibinfo {author} {\bibfnamefont {V.}~\bibnamefont
  {Paulisch}}, \bibinfo {author} {\bibfnamefont {H.~J.}\ \bibnamefont
  {Kimble}}, \bibinfo {author} {\bibfnamefont {J.~I.}\ \bibnamefont {Cirac}}, \
  and\ \bibinfo {author} {\bibfnamefont {A.}~\bibnamefont
  {Gonz\'alez-Tudela}},\ }\href {\doibase 10.1103/PhysRevA.97.053831}
  {\bibfield  {journal} {\bibinfo  {journal} {Phys. Rev. A}\ }\textbf {\bibinfo
  {volume} {97}},\ \bibinfo {pages} {053831} (\bibinfo {year}
  {2018})}\BibitemShut {NoStop}%
\bibitem [{\citenamefont {Quijandr\'{\i}a}\ \emph {et~al.}(2018)\citenamefont
  {Quijandr\'{\i}a}, \citenamefont {Strandberg},\ and\ \citenamefont
  {Johansson}}]{QuijJoha18}%
  \BibitemOpen
  \bibfield  {author} {\bibinfo {author} {\bibfnamefont {F.}~\bibnamefont
  {Quijandr\'{\i}a}}, \bibinfo {author} {\bibfnamefont {I.}~\bibnamefont
  {Strandberg}}, \ and\ \bibinfo {author} {\bibfnamefont {G.}~\bibnamefont
  {Johansson}},\ }\href {\doibase 10.1103/PhysRevLett.121.263603} {\bibfield
  {journal} {\bibinfo  {journal} {Phys. Rev. Lett.}\ }\textbf {\bibinfo
  {volume} {121}},\ \bibinfo {pages} {263603} (\bibinfo {year}
  {2018})}\BibitemShut {NoStop}%
\bibitem [{\citenamefont {Zheng}\ \emph {et~al.}(2010)\citenamefont {Zheng},
  \citenamefont {Gauthier},\ and\ \citenamefont {Baranger}}]{ZhenBara10}%
  \BibitemOpen
  \bibfield  {author} {\bibinfo {author} {\bibfnamefont {H.}~\bibnamefont
  {Zheng}}, \bibinfo {author} {\bibfnamefont {D.~J.}\ \bibnamefont {Gauthier}},
  \ and\ \bibinfo {author} {\bibfnamefont {H.~U.}\ \bibnamefont {Baranger}},\
  }\href {\doibase 10.1103/PhysRevA.82.063816} {\bibfield  {journal} {\bibinfo
  {journal} {Phys. Rev. A}\ }\textbf {\bibinfo {volume} {82}},\ \bibinfo
  {pages} {063816} (\bibinfo {year} {2010})}\BibitemShut {NoStop}%
\bibitem [{\citenamefont {Tufarelli}\ \emph {et~al.}(2014)\citenamefont
  {Tufarelli}, \citenamefont {Kim},\ and\ \citenamefont
  {Ciccarello}}]{Tufarelli2014}%
  \BibitemOpen
  \bibfield  {author} {\bibinfo {author} {\bibfnamefont {T.}~\bibnamefont
  {Tufarelli}}, \bibinfo {author} {\bibfnamefont {M.~S.}\ \bibnamefont {Kim}},
  \ and\ \bibinfo {author} {\bibfnamefont {F.}~\bibnamefont {Ciccarello}},\
  }\href {\doibase 10.1103/PhysRevA.90.012113} {\bibfield  {journal} {\bibinfo
  {journal} {Phys. Rev. A}\ }\textbf {\bibinfo {volume} {90}},\ \bibinfo
  {pages} {012113} (\bibinfo {year} {2014})}\BibitemShut {NoStop}%
\bibitem [{\citenamefont {Sanchez-Burillo}\ \emph {et~al.}(2014)\citenamefont
  {Sanchez-Burillo}, \citenamefont {Zueco}, \citenamefont {Garcia-Ripoll},\
  and\ \citenamefont {Martin-Moreno}}]{SancMart14}%
  \BibitemOpen
  \bibfield  {author} {\bibinfo {author} {\bibfnamefont {E.}~\bibnamefont
  {Sanchez-Burillo}}, \bibinfo {author} {\bibfnamefont {D.}~\bibnamefont
  {Zueco}}, \bibinfo {author} {\bibfnamefont {J.~J.}\ \bibnamefont
  {Garcia-Ripoll}}, \ and\ \bibinfo {author} {\bibfnamefont {L.}~\bibnamefont
  {Martin-Moreno}},\ }\href {\doibase 10.1103/PhysRevLett.113.263604}
  {\bibfield  {journal} {\bibinfo  {journal} {Phys. Rev. Lett.}\ }\textbf
  {\bibinfo {volume} {113}},\ \bibinfo {pages} {263604} (\bibinfo {year}
  {2014})}\BibitemShut {NoStop}%
\bibitem [{\citenamefont {S\'anchez-Burillo}\ \emph {et~al.}(2016)\citenamefont
  {S\'anchez-Burillo}, \citenamefont {Mart\'{\i}n-Moreno}, \citenamefont
  {Garc\'{\i}a-Ripoll},\ and\ \citenamefont {Zueco}}]{SancZuec16}%
  \BibitemOpen
  \bibfield  {author} {\bibinfo {author} {\bibfnamefont {E.}~\bibnamefont
  {S\'anchez-Burillo}}, \bibinfo {author} {\bibfnamefont {L.}~\bibnamefont
  {Mart\'{\i}n-Moreno}}, \bibinfo {author} {\bibfnamefont {J.~J.}\ \bibnamefont
  {Garc\'{\i}a-Ripoll}}, \ and\ \bibinfo {author} {\bibfnamefont
  {D.}~\bibnamefont {Zueco}},\ }\href {\doibase 10.1103/PhysRevA.94.053814}
  {\bibfield  {journal} {\bibinfo  {journal} {Phys. Rev. A}\ }\textbf {\bibinfo
  {volume} {94}},\ \bibinfo {pages} {053814} (\bibinfo {year}
  {2016})}\BibitemShut {NoStop}%
\bibitem [{\citenamefont {S\'anchez-Burillo}\ \emph {et~al.}(2017)\citenamefont
  {S\'anchez-Burillo}, \citenamefont {Zueco}, \citenamefont
  {Mart\'{\i}n-Moreno},\ and\ \citenamefont {Garc\'{\i}a-Ripoll}}]{SancGarc17}%
  \BibitemOpen
  \bibfield  {author} {\bibinfo {author} {\bibfnamefont {E.}~\bibnamefont
  {S\'anchez-Burillo}}, \bibinfo {author} {\bibfnamefont {D.}~\bibnamefont
  {Zueco}}, \bibinfo {author} {\bibfnamefont {L.}~\bibnamefont
  {Mart\'{\i}n-Moreno}}, \ and\ \bibinfo {author} {\bibfnamefont {J.~J.}\
  \bibnamefont {Garc\'{\i}a-Ripoll}},\ }\href {\doibase
  10.1103/PhysRevA.96.023831} {\bibfield  {journal} {\bibinfo  {journal} {Phys.
  Rev. A}\ }\textbf {\bibinfo {volume} {96}},\ \bibinfo {pages} {023831}
  (\bibinfo {year} {2017})}\BibitemShut {NoStop}%
\bibitem [{\citenamefont {Wallraff}\ \emph {et~al.}(2004)\citenamefont
  {Wallraff}, \citenamefont {Schuster}, \citenamefont {Blais}, \citenamefont
  {Frunzio}, \citenamefont {Huang}, \citenamefont {Majer}, \citenamefont
  {Kumar}, \citenamefont {Girvin},\ and\ \citenamefont
  {Schoelkopf}}]{Wallraff2004}%
  \BibitemOpen
  \bibfield  {author} {\bibinfo {author} {\bibfnamefont {A.}~\bibnamefont
  {Wallraff}}, \bibinfo {author} {\bibfnamefont {D.~I.}\ \bibnamefont
  {Schuster}}, \bibinfo {author} {\bibfnamefont {A.}~\bibnamefont {Blais}},
  \bibinfo {author} {\bibfnamefont {L.}~\bibnamefont {Frunzio}}, \bibinfo
  {author} {\bibfnamefont {R.-S.}\ \bibnamefont {Huang}}, \bibinfo {author}
  {\bibfnamefont {J.}~\bibnamefont {Majer}}, \bibinfo {author} {\bibfnamefont
  {S.}~\bibnamefont {Kumar}}, \bibinfo {author} {\bibfnamefont {S.~M.}\
  \bibnamefont {Girvin}}, \ and\ \bibinfo {author} {\bibfnamefont {R.~J.}\
  \bibnamefont {Schoelkopf}},\ }\href {https://doi.org/10.1038/nature02851
  http://10.0.4.14/nature02851} {\bibfield  {journal} {\bibinfo  {journal}
  {Nature}\ }\textbf {\bibinfo {volume} {431}},\ \bibinfo {pages} {162}
  (\bibinfo {year} {2004})}\BibitemShut {NoStop}%
\bibitem [{\citenamefont {Blais}\ \emph {et~al.}(2004)\citenamefont {Blais},
  \citenamefont {Huang}, \citenamefont {Wallraff}, \citenamefont {Girvin},\
  and\ \citenamefont {Schoelkopf}}]{Blais2004}%
  \BibitemOpen
  \bibfield  {author} {\bibinfo {author} {\bibfnamefont {A.}~\bibnamefont
  {Blais}}, \bibinfo {author} {\bibfnamefont {R.~S.}\ \bibnamefont {Huang}},
  \bibinfo {author} {\bibfnamefont {A.}~\bibnamefont {Wallraff}}, \bibinfo
  {author} {\bibfnamefont {S.~M.}\ \bibnamefont {Girvin}}, \ and\ \bibinfo
  {author} {\bibfnamefont {R.~J.}\ \bibnamefont {Schoelkopf}},\ }\href
  {\doibase 10.1103/PhysRevA.69.062320} {\bibfield  {journal} {\bibinfo
  {journal} {Physical Review A - Atomic, Molecular, and Optical Physics}\
  }\textbf {\bibinfo {volume} {69}},\ \bibinfo {pages} {1} (\bibinfo {year}
  {2004})},\ \Eprint {http://arxiv.org/abs/0402216} {arXiv:0402216 [cond-mat]}
  \BibitemShut {NoStop}%
\bibitem [{\citenamefont {Wendin}(2017)}]{Wendin2017}%
  \BibitemOpen
  \bibfield  {author} {\bibinfo {author} {\bibfnamefont {G.}~\bibnamefont
  {Wendin}},\ }\href {\doibase 10.1088/1361-6633/aa7e1a} {\bibfield  {journal}
  {\bibinfo  {journal} {Reports on Progress in Physics}\ }\textbf {\bibinfo
  {volume} {80}},\ \bibinfo {pages} {106001} (\bibinfo {year}
  {2017})}\BibitemShut {NoStop}%
\bibitem [{\citenamefont {F.~Kockum}\ and\ \citenamefont
  {Nori}(2019)}]{Kockum2019}%
  \BibitemOpen
  \bibfield  {author} {\bibinfo {author} {\bibfnamefont {A.}~\bibnamefont
  {F.~Kockum}}\ and\ \bibinfo {author} {\bibfnamefont {F.}~\bibnamefont
  {Nori}},\ }\href {arXiv:1908.09558v1} {\bibfield  {journal} {\bibinfo
  {journal} {arXiv:1908.09558v1}\ } (\bibinfo {year} {2019})}\BibitemShut
  {NoStop}%
\bibitem [{\citenamefont {Devoret}\ \emph {et~al.}(2007)\citenamefont
  {Devoret}, \citenamefont {Girvin},\ and\ \citenamefont
  {Schoelkopf}}]{DevoScho07}%
  \BibitemOpen
  \bibfield  {author} {\bibinfo {author} {\bibfnamefont {M.}~\bibnamefont
  {Devoret}}, \bibinfo {author} {\bibfnamefont {S.}~\bibnamefont {Girvin}}, \
  and\ \bibinfo {author} {\bibfnamefont {R.}~\bibnamefont {Schoelkopf}},\
  }\href {\doibase 10.1002/andp.200710261} {\bibfield  {journal} {\bibinfo
  {journal} {Annalen der Physik}\ }\textbf {\bibinfo {volume} {16}},\ \bibinfo
  {pages} {767} (\bibinfo {year} {2007})}\BibitemShut {NoStop}%
\bibitem [{\citenamefont {Bamba}\ and\ \citenamefont
  {Ogawa}(2014)}]{BambOgaw14}%
  \BibitemOpen
  \bibfield  {author} {\bibinfo {author} {\bibfnamefont {M.}~\bibnamefont
  {Bamba}}\ and\ \bibinfo {author} {\bibfnamefont {T.}~\bibnamefont {Ogawa}},\
  }\href {\doibase 10.1103/PhysRevA.89.023817} {\bibfield  {journal} {\bibinfo
  {journal} {Phys. Rev. A}\ }\textbf {\bibinfo {volume} {89}},\ \bibinfo
  {pages} {023817} (\bibinfo {year} {2014})}\BibitemShut {NoStop}%
\bibitem [{\citenamefont {Frisk~Kockum}\ \emph {et~al.}(2019)\citenamefont
  {Frisk~Kockum}, \citenamefont {Miranowicz}, \citenamefont {De~Liberato},
  \citenamefont {Savasta},\ and\ \citenamefont {Nori}}]{USC_Anton}%
  \BibitemOpen
  \bibfield  {author} {\bibinfo {author} {\bibfnamefont {A.}~\bibnamefont
  {Frisk~Kockum}}, \bibinfo {author} {\bibfnamefont {A.}~\bibnamefont
  {Miranowicz}}, \bibinfo {author} {\bibfnamefont {S.}~\bibnamefont
  {De~Liberato}}, \bibinfo {author} {\bibfnamefont {S.}~\bibnamefont
  {Savasta}}, \ and\ \bibinfo {author} {\bibfnamefont {F.}~\bibnamefont
  {Nori}},\ }\href {\doibase 10.1038/s42254-018-0006-2} {\bibfield  {journal}
  {\bibinfo  {journal} {Nature Reviews Physics}\ }\textbf {\bibinfo {volume}
  {1}},\ \bibinfo {pages} {19} (\bibinfo {year} {2019})}\BibitemShut {NoStop}%
\bibitem [{\citenamefont {Zueco}\ and\ \citenamefont
  {Garc\'{\i}a-Ripoll}(2019)}]{ZuecGarc19}%
  \BibitemOpen
  \bibfield  {author} {\bibinfo {author} {\bibfnamefont {D.}~\bibnamefont
  {Zueco}}\ and\ \bibinfo {author} {\bibfnamefont {J.}~\bibnamefont
  {Garc\'{\i}a-Ripoll}},\ }\href {\doibase 10.1103/PhysRevA.99.013807}
  {\bibfield  {journal} {\bibinfo  {journal} {Phys. Rev. A}\ }\textbf {\bibinfo
  {volume} {99}},\ \bibinfo {pages} {013807} (\bibinfo {year}
  {2019})}\BibitemShut {NoStop}%
\bibitem [{\citenamefont {Mlynek}\ \emph {et~al.}(2014)\citenamefont {Mlynek},
  \citenamefont {Abdumalikov}, \citenamefont {Eichler},\ and\ \citenamefont
  {Wallraff}}]{MlynWall14}%
  \BibitemOpen
  \bibfield  {author} {\bibinfo {author} {\bibfnamefont {J.~A.}\ \bibnamefont
  {Mlynek}}, \bibinfo {author} {\bibfnamefont {A.~A.}\ \bibnamefont
  {Abdumalikov}}, \bibinfo {author} {\bibfnamefont {C.}~\bibnamefont
  {Eichler}}, \ and\ \bibinfo {author} {\bibfnamefont {A.}~\bibnamefont
  {Wallraff}},\ }\href {https://doi.org/10.1038/ncomms6186} {\bibfield
  {journal} {\bibinfo  {journal} {Nature Communications}\ }\textbf {\bibinfo
  {volume} {5}},\ \bibinfo {pages} {5186 EP } (\bibinfo {year}
  {2014})}\BibitemShut {NoStop}%
\bibitem [{\citenamefont {Nataf}\ and\ \citenamefont
  {Ciuti}(2010)}]{NataCiut10}%
  \BibitemOpen
  \bibfield  {author} {\bibinfo {author} {\bibfnamefont {P.}~\bibnamefont
  {Nataf}}\ and\ \bibinfo {author} {\bibfnamefont {C.}~\bibnamefont {Ciuti}},\
  }\href {https://doi.org/10.1038/ncomms1069} {\bibfield  {journal} {\bibinfo
  {journal} {Nature Communications}\ }\textbf {\bibinfo {volume} {1}},\
  \bibinfo {pages} {72 EP } (\bibinfo {year} {2010})}\BibitemShut {NoStop}%
\bibitem [{\citenamefont {Bamba}\ \emph {et~al.}(2016)\citenamefont {Bamba},
  \citenamefont {Inomata},\ and\ \citenamefont {Nakamura}}]{BambNaka16}%
  \BibitemOpen
  \bibfield  {author} {\bibinfo {author} {\bibfnamefont {M.}~\bibnamefont
  {Bamba}}, \bibinfo {author} {\bibfnamefont {K.}~\bibnamefont {Inomata}}, \
  and\ \bibinfo {author} {\bibfnamefont {Y.}~\bibnamefont {Nakamura}},\ }\href
  {\doibase 10.1103/PhysRevLett.117.173601} {\bibfield  {journal} {\bibinfo
  {journal} {Phys. Rev. Lett.}\ }\textbf {\bibinfo {volume} {117}},\ \bibinfo
  {pages} {173601} (\bibinfo {year} {2016})}\BibitemShut {NoStop}%
\bibitem [{\citenamefont {Bamba}\ and\ \citenamefont
  {Imoto}(2017)}]{BambNobu17}%
  \BibitemOpen
  \bibfield  {author} {\bibinfo {author} {\bibfnamefont {M.}~\bibnamefont
  {Bamba}}\ and\ \bibinfo {author} {\bibfnamefont {N.}~\bibnamefont {Imoto}},\
  }\href {\doibase 10.1103/PhysRevA.96.053857} {\bibfield  {journal} {\bibinfo
  {journal} {Phys. Rev. A}\ }\textbf {\bibinfo {volume} {96}},\ \bibinfo
  {pages} {053857} (\bibinfo {year} {2017})}\BibitemShut {NoStop}%
\bibitem [{\citenamefont {Hoi}\ \emph {et~al.}(2015)\citenamefont {Hoi},
  \citenamefont {Kockum}, \citenamefont {Tornberg}, \citenamefont
  {Pourkabirian}, \citenamefont {Johansson}, \citenamefont {Delsing},\ and\
  \citenamefont {Wilson}}]{Hoi2015}%
  \BibitemOpen
  \bibfield  {author} {\bibinfo {author} {\bibfnamefont {I.~C.}\ \bibnamefont
  {Hoi}}, \bibinfo {author} {\bibfnamefont {A.~F.}\ \bibnamefont {Kockum}},
  \bibinfo {author} {\bibfnamefont {L.}~\bibnamefont {Tornberg}}, \bibinfo
  {author} {\bibfnamefont {A.}~\bibnamefont {Pourkabirian}}, \bibinfo {author}
  {\bibfnamefont {G.}~\bibnamefont {Johansson}}, \bibinfo {author}
  {\bibfnamefont {P.}~\bibnamefont {Delsing}}, \ and\ \bibinfo {author}
  {\bibfnamefont {C.~M.}\ \bibnamefont {Wilson}},\ }\href {\doibase
  10.1038/nphys3484} {\bibfield  {journal} {\bibinfo  {journal} {Nature
  Physics}\ }\textbf {\bibinfo {volume} {11}},\ \bibinfo {pages} {1045}
  (\bibinfo {year} {2015})},\ \Eprint {http://arxiv.org/abs/1410.8840}
  {arXiv:1410.8840} \BibitemShut {NoStop}%
\bibitem [{\citenamefont {Koch}\ \emph {et~al.}(2007)\citenamefont {Koch},
  \citenamefont {Yu}, \citenamefont {Gambetta}, \citenamefont {Houck},
  \citenamefont {Schuster}, \citenamefont {Majer}, \citenamefont {Blais},
  \citenamefont {Devoret}, \citenamefont {Girvin},\ and\ \citenamefont
  {Schoelkopf}}]{Koch2007}%
  \BibitemOpen
  \bibfield  {author} {\bibinfo {author} {\bibfnamefont {J.}~\bibnamefont
  {Koch}}, \bibinfo {author} {\bibfnamefont {T.~M.}\ \bibnamefont {Yu}},
  \bibinfo {author} {\bibfnamefont {J.}~\bibnamefont {Gambetta}}, \bibinfo
  {author} {\bibfnamefont {A.~A.}\ \bibnamefont {Houck}}, \bibinfo {author}
  {\bibfnamefont {D.~I.}\ \bibnamefont {Schuster}}, \bibinfo {author}
  {\bibfnamefont {J.}~\bibnamefont {Majer}}, \bibinfo {author} {\bibfnamefont
  {A.}~\bibnamefont {Blais}}, \bibinfo {author} {\bibfnamefont {M.~H.}\
  \bibnamefont {Devoret}}, \bibinfo {author} {\bibfnamefont {S.~M.}\
  \bibnamefont {Girvin}}, \ and\ \bibinfo {author} {\bibfnamefont {R.~J.}\
  \bibnamefont {Schoelkopf}},\ }\href {\doibase 10.1103/PhysRevA.76.042319}
  {\bibfield  {journal} {\bibinfo  {journal} {Physical Review A - Atomic,
  Molecular, and Optical Physics}\ }\textbf {\bibinfo {volume} {76}},\ \bibinfo
  {pages} {1} (\bibinfo {year} {2007})},\ \Eprint
  {http://arxiv.org/abs/0703002} {arXiv:0703002 [cond-mat]} \BibitemShut
  {NoStop}%
\bibitem [{\citenamefont {Wen}\ \emph {et~al.}(2018)\citenamefont {Wen},
  \citenamefont {Kockum}, \citenamefont {Ian}, \citenamefont {Chen},
  \citenamefont {Nori},\ and\ \citenamefont {Hoi}}]{Wen2018}%
  \BibitemOpen
  \bibfield  {author} {\bibinfo {author} {\bibfnamefont {P.~Y.}\ \bibnamefont
  {Wen}}, \bibinfo {author} {\bibfnamefont {A.~F.}\ \bibnamefont {Kockum}},
  \bibinfo {author} {\bibfnamefont {H.}~\bibnamefont {Ian}}, \bibinfo {author}
  {\bibfnamefont {J.~C.}\ \bibnamefont {Chen}}, \bibinfo {author}
  {\bibfnamefont {F.}~\bibnamefont {Nori}}, \ and\ \bibinfo {author}
  {\bibfnamefont {I.-C.}\ \bibnamefont {Hoi}},\ }\href {\doibase
  10.1103/PhysRevLett.120.063603} {\bibfield  {journal} {\bibinfo  {journal}
  {Phys. Rev. Lett.}\ }\textbf {\bibinfo {volume} {120}},\ \bibinfo {pages}
  {063603} (\bibinfo {year} {2018})}\BibitemShut {NoStop}%
\bibitem [{\citenamefont {Wen}\ \emph {et~al.}(2019)\citenamefont {Wen},
  \citenamefont {Lin}, \citenamefont {F.~Kockum}, \citenamefont {Suri},
  \citenamefont {Ian},\ and\ \citenamefont {Chen}}]{Wen2019}%
  \BibitemOpen
  \bibfield  {author} {\bibinfo {author} {\bibfnamefont {P.~Y.}\ \bibnamefont
  {Wen}}, \bibinfo {author} {\bibfnamefont {K.-T.}\ \bibnamefont {Lin}},
  \bibinfo {author} {\bibfnamefont {A.}~\bibnamefont {F.~Kockum}}, \bibinfo
  {author} {\bibfnamefont {B.}~\bibnamefont {Suri}}, \bibinfo {author}
  {\bibfnamefont {H.}~\bibnamefont {Ian}}, \ and\ \bibinfo {author}
  {\bibfnamefont {J.~a.}\ \bibnamefont {Chen}},\ }\href {arXiv:1904.12473v1}
  {\bibfield  {journal} {\bibinfo  {journal} {arXiv:1904.12473v1}\ } (\bibinfo
  {year} {2019})}\BibitemShut {NoStop}%
\bibitem [{\citenamefont {Peng}\ \emph {et~al.}(2016)\citenamefont {Peng},
  \citenamefont {de~Graaf}, \citenamefont {Tsai},\ and\ \citenamefont
  {Astafiev}}]{Peng2016}%
  \BibitemOpen
  \bibfield  {author} {\bibinfo {author} {\bibfnamefont {Z.~H.}\ \bibnamefont
  {Peng}}, \bibinfo {author} {\bibfnamefont {S.~E.}\ \bibnamefont {de~Graaf}},
  \bibinfo {author} {\bibfnamefont {J.~S.}\ \bibnamefont {Tsai}}, \ and\
  \bibinfo {author} {\bibfnamefont {O.~V.}\ \bibnamefont {Astafiev}},\ }\href
  {\doibase 10.1038/ncomms12588} {\bibfield  {journal} {\bibinfo  {journal}
  {Nature Communications}\ }\textbf {\bibinfo {volume} {7}},\ \bibinfo {pages}
  {12588} (\bibinfo {year} {2016})}\BibitemShut {NoStop}%
\bibitem [{\citenamefont {Forn-D\'{\i}az}\ \emph {et~al.}(2017)\citenamefont
  {Forn-D\'{\i}az}, \citenamefont {Warren}, \citenamefont {Chang},
  \citenamefont {Vadiraj},\ and\ \citenamefont {Wilson}}]{FornDiaz2017}%
  \BibitemOpen
  \bibfield  {author} {\bibinfo {author} {\bibfnamefont {P.}~\bibnamefont
  {Forn-D\'{\i}az}}, \bibinfo {author} {\bibfnamefont {C.~W.}\ \bibnamefont
  {Warren}}, \bibinfo {author} {\bibfnamefont {C.~W.~S.}\ \bibnamefont
  {Chang}}, \bibinfo {author} {\bibfnamefont {A.~M.}\ \bibnamefont {Vadiraj}},
  \ and\ \bibinfo {author} {\bibfnamefont {C.~M.}\ \bibnamefont {Wilson}},\
  }\href {\doibase 10.1103/PhysRevApplied.8.054015} {\bibfield  {journal}
  {\bibinfo  {journal} {Phys. Rev. Applied}\ }\textbf {\bibinfo {volume} {8}},\
  \bibinfo {pages} {054015} (\bibinfo {year} {2017})}\BibitemShut {NoStop}%
\bibitem [{\citenamefont {Dorner}\ and\ \citenamefont
  {Zoller}(2002)}]{Dorner2002}%
  \BibitemOpen
  \bibfield  {author} {\bibinfo {author} {\bibfnamefont {U.}~\bibnamefont
  {Dorner}}\ and\ \bibinfo {author} {\bibfnamefont {P.}~\bibnamefont
  {Zoller}},\ }\href {\doibase 10.1103/PhysRevA.66.023816} {\bibfield
  {journal} {\bibinfo  {journal} {Physical Review A - Atomic, Molecular, and
  Optical Physics}\ }\textbf {\bibinfo {volume} {66}},\ \bibinfo {pages} {1}
  (\bibinfo {year} {2002})},\ \Eprint {http://arxiv.org/abs/0203147}
  {arXiv:0203147 [quant-ph]} \BibitemShut {NoStop}%
\bibitem [{\citenamefont {Guo}\ \emph {et~al.}(2017)\citenamefont {Guo},
  \citenamefont {Grimsmo}, \citenamefont {Kockum}, \citenamefont {Pletyukhov},\
  and\ \citenamefont {Johansson}}]{Guo2017}%
  \BibitemOpen
  \bibfield  {author} {\bibinfo {author} {\bibfnamefont {L.}~\bibnamefont
  {Guo}}, \bibinfo {author} {\bibfnamefont {A.}~\bibnamefont {Grimsmo}},
  \bibinfo {author} {\bibfnamefont {A.~F.}\ \bibnamefont {Kockum}}, \bibinfo
  {author} {\bibfnamefont {M.}~\bibnamefont {Pletyukhov}}, \ and\ \bibinfo
  {author} {\bibfnamefont {G.}~\bibnamefont {Johansson}},\ }\href {\doibase
  10.1103/PhysRevA.95.053821} {\bibfield  {journal} {\bibinfo  {journal} {Phys.
  Rev. A}\ }\textbf {\bibinfo {volume} {95}},\ \bibinfo {pages} {053821}
  (\bibinfo {year} {2017})}\BibitemShut {NoStop}%
\bibitem [{\citenamefont {Grimsmo}(2015)}]{Grimsmo2015}%
  \BibitemOpen
  \bibfield  {author} {\bibinfo {author} {\bibfnamefont {A.~L.}\ \bibnamefont
  {Grimsmo}},\ }\href {\doibase 10.1103/PhysRevLett.115.060402} {\bibfield
  {journal} {\bibinfo  {journal} {Phys. Rev. Lett.}\ }\textbf {\bibinfo
  {volume} {115}},\ \bibinfo {pages} {060402} (\bibinfo {year}
  {2015})}\BibitemShut {NoStop}%
\bibitem [{\citenamefont {Pichler}\ and\ \citenamefont
  {Zoller}(2016)}]{Pichler2016}%
  \BibitemOpen
  \bibfield  {author} {\bibinfo {author} {\bibfnamefont {H.}~\bibnamefont
  {Pichler}}\ and\ \bibinfo {author} {\bibfnamefont {P.}~\bibnamefont
  {Zoller}},\ }\href {\doibase 10.1103/PhysRevLett.116.093601} {\bibfield
  {journal} {\bibinfo  {journal} {Phys. Rev. Lett.}\ }\textbf {\bibinfo
  {volume} {116}},\ \bibinfo {pages} {093601} (\bibinfo {year}
  {2016})}\BibitemShut {NoStop}%
\bibitem [{\citenamefont {Pichler}\ \emph {et~al.}(2017)\citenamefont
  {Pichler}, \citenamefont {Choi}, \citenamefont {Zoller},\ and\ \citenamefont
  {Lukin}}]{Pichler2017}%
  \BibitemOpen
  \bibfield  {author} {\bibinfo {author} {\bibfnamefont {H.}~\bibnamefont
  {Pichler}}, \bibinfo {author} {\bibfnamefont {S.}~\bibnamefont {Choi}},
  \bibinfo {author} {\bibfnamefont {P.}~\bibnamefont {Zoller}}, \ and\ \bibinfo
  {author} {\bibfnamefont {M.~D.}\ \bibnamefont {Lukin}},\ }\href {\doibase
  10.1073/pnas.1711003114} {\bibfield  {journal} {\bibinfo  {journal}
  {Proceedings of the National Academy of Sciences}\ }\textbf {\bibinfo
  {volume} {114}},\ \bibinfo {pages} {11362} (\bibinfo {year} {2017})},\
  \Eprint
  {http://arxiv.org/abs/https://www.pnas.org/content/114/43/11362.full.pdf}
  {https://www.pnas.org/content/114/43/11362.full.pdf} \BibitemShut {NoStop}%
\bibitem [{\citenamefont {Guimond}\ \emph {et~al.}(2017)\citenamefont
  {Guimond}, \citenamefont {Pletyukhov}, \citenamefont {Pichler},\ and\
  \citenamefont {Zoller}}]{Guimond_2017}%
  \BibitemOpen
  \bibfield  {author} {\bibinfo {author} {\bibfnamefont {P.-O.}\ \bibnamefont
  {Guimond}}, \bibinfo {author} {\bibfnamefont {M.}~\bibnamefont {Pletyukhov}},
  \bibinfo {author} {\bibfnamefont {H.}~\bibnamefont {Pichler}}, \ and\
  \bibinfo {author} {\bibfnamefont {P.}~\bibnamefont {Zoller}},\ }\href
  {\doibase 10.1088/2058-9565/aa7f03} {\bibfield  {journal} {\bibinfo
  {journal} {Quantum Science and Technology}\ }\textbf {\bibinfo {volume}
  {2}},\ \bibinfo {pages} {044012} (\bibinfo {year} {2017})}\BibitemShut
  {NoStop}%
\bibitem [{\citenamefont {Schneider}\ \emph {et~al.}(2016)\citenamefont
  {Schneider}, \citenamefont {Sproll}, \citenamefont {Stawiarski},
  \citenamefont {Schmitteckert},\ and\ \citenamefont {Busch}}]{SchnBusc16}%
  \BibitemOpen
  \bibfield  {author} {\bibinfo {author} {\bibfnamefont {M.~P.}\ \bibnamefont
  {Schneider}}, \bibinfo {author} {\bibfnamefont {T.}~\bibnamefont {Sproll}},
  \bibinfo {author} {\bibfnamefont {C.}~\bibnamefont {Stawiarski}}, \bibinfo
  {author} {\bibfnamefont {P.}~\bibnamefont {Schmitteckert}}, \ and\ \bibinfo
  {author} {\bibfnamefont {K.}~\bibnamefont {Busch}},\ }\href {\doibase
  10.1103/PhysRevA.93.013828} {\bibfield  {journal} {\bibinfo  {journal} {Phys.
  Rev. A}\ }\textbf {\bibinfo {volume} {93}},\ \bibinfo {pages} {013828}
  (\bibinfo {year} {2016})}\BibitemShut {NoStop}%
\bibitem [{\citenamefont {Gonz\'alez-Tudela}\ and\ \citenamefont
  {Cirac}(2017)}]{GonzCira17}%
  \BibitemOpen
  \bibfield  {author} {\bibinfo {author} {\bibfnamefont {A.}~\bibnamefont
  {Gonz\'alez-Tudela}}\ and\ \bibinfo {author} {\bibfnamefont {J.~I.}\
  \bibnamefont {Cirac}},\ }\href {\doibase 10.1103/PhysRevA.96.043811}
  {\bibfield  {journal} {\bibinfo  {journal} {Phys. Rev. A}\ }\textbf {\bibinfo
  {volume} {96}},\ \bibinfo {pages} {043811} (\bibinfo {year}
  {2017})}\BibitemShut {NoStop}%
\bibitem [{\citenamefont {Devoret}(1995)}]{Devoret1995}%
  \BibitemOpen
  \bibfield  {author} {\bibinfo {author} {\bibfnamefont {M.~H.}\ \bibnamefont
  {Devoret}},\ }\href@noop {} {\bibfield  {journal} {\bibinfo  {journal} {Les
  Houches Session LXIII}\ ,\ \bibinfo {pages} {p. 351}} (\bibinfo {year}
  {1995})}\BibitemShut {NoStop}%
\bibitem [{\citenamefont {Wei\ss{}l}\ \emph {et~al.}(2015)\citenamefont
  {Wei\ss{}l}, \citenamefont {K\"ung}, \citenamefont {Dumur}, \citenamefont
  {Feofanov}, \citenamefont {Matei}, \citenamefont {Naud}, \citenamefont
  {Buisson}, \citenamefont {Hekking},\ and\ \citenamefont
  {Guichard}}]{Weissl2015}%
  \BibitemOpen
  \bibfield  {author} {\bibinfo {author} {\bibfnamefont {T.}~\bibnamefont
  {Wei\ss{}l}}, \bibinfo {author} {\bibfnamefont {B.}~\bibnamefont {K\"ung}},
  \bibinfo {author} {\bibfnamefont {E.}~\bibnamefont {Dumur}}, \bibinfo
  {author} {\bibfnamefont {A.~K.}\ \bibnamefont {Feofanov}}, \bibinfo {author}
  {\bibfnamefont {I.}~\bibnamefont {Matei}}, \bibinfo {author} {\bibfnamefont
  {C.}~\bibnamefont {Naud}}, \bibinfo {author} {\bibfnamefont {O.}~\bibnamefont
  {Buisson}}, \bibinfo {author} {\bibfnamefont {F.~W.~J.}\ \bibnamefont
  {Hekking}}, \ and\ \bibinfo {author} {\bibfnamefont {W.}~\bibnamefont
  {Guichard}},\ }\href {\doibase 10.1103/PhysRevB.92.104508} {\bibfield
  {journal} {\bibinfo  {journal} {Phys. Rev. B}\ }\textbf {\bibinfo {volume}
  {92}},\ \bibinfo {pages} {104508} (\bibinfo {year} {2015})}\BibitemShut
  {NoStop}%
\bibitem [{\citenamefont {Krupko}\ \emph {et~al.}(2018)\citenamefont {Krupko},
  \citenamefont {Nguyen}, \citenamefont {Wei\ss{}l}, \citenamefont {Dumur},
  \citenamefont {Puertas}, \citenamefont {Dassonneville}, \citenamefont {Naud},
  \citenamefont {Hekking}, \citenamefont {Basko}, \citenamefont {Buisson},
  \citenamefont {Roch},\ and\ \citenamefont {Hasch-Guichard}}]{Krupko2018}%
  \BibitemOpen
  \bibfield  {author} {\bibinfo {author} {\bibfnamefont {Y.}~\bibnamefont
  {Krupko}}, \bibinfo {author} {\bibfnamefont {V.~D.}\ \bibnamefont {Nguyen}},
  \bibinfo {author} {\bibfnamefont {T.}~\bibnamefont {Wei\ss{}l}}, \bibinfo
  {author} {\bibfnamefont {E.}~\bibnamefont {Dumur}}, \bibinfo {author}
  {\bibfnamefont {J.}~\bibnamefont {Puertas}}, \bibinfo {author} {\bibfnamefont
  {R.}~\bibnamefont {Dassonneville}}, \bibinfo {author} {\bibfnamefont
  {C.}~\bibnamefont {Naud}}, \bibinfo {author} {\bibfnamefont {F.~W.~J.}\
  \bibnamefont {Hekking}}, \bibinfo {author} {\bibfnamefont {D.~M.}\
  \bibnamefont {Basko}}, \bibinfo {author} {\bibfnamefont {O.}~\bibnamefont
  {Buisson}}, \bibinfo {author} {\bibfnamefont {N.}~\bibnamefont {Roch}}, \
  and\ \bibinfo {author} {\bibfnamefont {W.}~\bibnamefont {Hasch-Guichard}},\
  }\href {\doibase 10.1103/PhysRevB.98.094516} {\bibfield  {journal} {\bibinfo
  {journal} {Phys. Rev. B}\ }\textbf {\bibinfo {volume} {98}},\ \bibinfo
  {pages} {094516} (\bibinfo {year} {2018})}\BibitemShut {NoStop}%
\bibitem [{\citenamefont {Masluk}\ \emph {et~al.}(2012)\citenamefont {Masluk},
  \citenamefont {Pop}, \citenamefont {Kamal}, \citenamefont {Minev},\ and\
  \citenamefont {Devoret}}]{Masluk2012}%
  \BibitemOpen
  \bibfield  {author} {\bibinfo {author} {\bibfnamefont {N.~A.}\ \bibnamefont
  {Masluk}}, \bibinfo {author} {\bibfnamefont {I.~M.}\ \bibnamefont {Pop}},
  \bibinfo {author} {\bibfnamefont {A.}~\bibnamefont {Kamal}}, \bibinfo
  {author} {\bibfnamefont {Z.~K.}\ \bibnamefont {Minev}}, \ and\ \bibinfo
  {author} {\bibfnamefont {M.~H.}\ \bibnamefont {Devoret}},\ }\href {\doibase
  10.1103/PhysRevLett.109.137002} {\bibfield  {journal} {\bibinfo  {journal}
  {Phys. Rev. Lett.}\ }\textbf {\bibinfo {volume} {109}},\ \bibinfo {pages}
  {137002} (\bibinfo {year} {2012})}\BibitemShut {NoStop}%
\bibitem [{\citenamefont {Purcell}(1946)}]{Purcell}%
  \BibitemOpen
  \bibfield  {author} {\bibinfo {author} {\bibfnamefont {E.~M.}\ \bibnamefont
  {Purcell}},\ }\href@noop {} {\bibfield  {journal} {\bibinfo  {journal} {Phys.
  Rev.}\ }\textbf {\bibinfo {volume} {69}},\ \bibinfo {pages} {681} (\bibinfo
  {year} {1946})}\BibitemShut {NoStop}%
\bibitem [{\citenamefont {Tufarelli}\ \emph {et~al.}(2013)\citenamefont
  {Tufarelli}, \citenamefont {Ciccarello},\ and\ \citenamefont
  {Kim}}]{TufaKim13}%
  \BibitemOpen
  \bibfield  {author} {\bibinfo {author} {\bibfnamefont {T.}~\bibnamefont
  {Tufarelli}}, \bibinfo {author} {\bibfnamefont {F.}~\bibnamefont
  {Ciccarello}}, \ and\ \bibinfo {author} {\bibfnamefont {M.~S.}\ \bibnamefont
  {Kim}},\ }\href {\doibase 10.1103/PhysRevA.87.013820} {\bibfield  {journal}
  {\bibinfo  {journal} {Phys. Rev. A}\ }\textbf {\bibinfo {volume} {87}},\
  \bibinfo {pages} {013820} (\bibinfo {year} {2013})}\BibitemShut {NoStop}%
\end{thebibliography}%

	%
	%
	%
	%
	%
	%
	%
	%
	%

\end{document}